\documentclass[aps,prd,twocolumn,showpacs,eqsecnum,floatfix,amssymb,amsmath,showpacs,a4paper,superscriptaddress,nofootinbib]{revtex4-1}
\usepackage{graphicx,multirow}
\usepackage{amsfonts}
\usepackage{units}
\usepackage[latin1]{inputenc}      		% tradução de acentos
\usepackage{dcolumn}
\usepackage{array}
\usepackage{siunitx}

\def\be{\begin{equation}}
\def\ee{\end{equation}}
\def\ba{\begin{align}}
\def\ea{\end{align}}

\def\X{\multirow{2}{*}{X}}
\begin{document}
%--------------------------------------------------------------
\setcounter{topnumber}{1}

\title{Quasinormal modes of extremal black holes}
%--------------------------------------------------------------
\author{Maur\'icio Richartz}
\email{mauricio.richartz@ufabc.edu.br}
\affiliation{Centro de Matem\'atica, Computa\c{c}\~ao e Cogni\c{c}\~ao, Universidade Federal do ABC (UFABC), 09210-170 Santo Andr\'e, SP, Brazil}
%--------------------------------------------------------------
%------------------------------------------------------------------------------------------------------------------------------------------
%------------------------------------------------------------------------------------------------------------------------------------------
%------------------------------------------------------------------------------------------------------------------------------------------
%------------------------------------------------------------------------------------------------------------------------------------------
%
%------------------------------------------------------------------------------------------------------------------------------------------

%--------------------------------------------------------------
\begin{abstract}
%--------------------------------------------------------------

The continued fraction method (also known as Leaver's method) is one of the most effective techniques used to determine the quasinormal modes of a black hole. For extremal black holes, however, the method does not work (since, in such a case, the event horizon is an irregular singular point of the associated wave equation). Fortunately, there exists a modified version of the method, devised by Onozawa \emph{et al.}~[Phys.~Rev.~D {\bf{53}}, 7033 (1996)], which works for neutral massless fields around an extremal Reissner-Nordstr\"om black hole. In this paper, we generalize the ideas of Onozawa \emph{et al.}~to charged massless perturbations around an extremal Reissner-Nordstr\"om black hole and to neutral massless perturbations around an extremal Kerr black hole. In particular, the existence of damped modes is analysed in detail. Similarities and differences between the results of the original continued fraction method for near extremal black holes and the results of the new continued fraction method for extremal black holes are discussed. Mode stability of extremal black holes is also investigated. 
%--------------------------------------------------------------
\end{abstract}
%--------------------------------------------------------------
\maketitle

%================
\section{Introduction}
%================

Perturbation theory is used everywhere in Physics, from the three-body problem of Celestial Mechanics to Feynman diagrams of Quantum Field Theory. In General Relativity, in particular, linear perturbations of a background metric (or, similarly, linear perturbations of a spin-s field in a curved spacetime) satisfy a second order linear PDE. In some cases, e.g.~if the spacetime is spherically symmetric (as in a Schwarzschild black hole) or, more generally, Petrov-type D (as in a Kerr black hole), the PDE can be transformed, by separation of variables, into a system of ODEs which is more amenable to mathematical treatment.

Historically, Regge and Wheeler, in the late 1950s, were the first to employ perturbation theory in Black Hole Physics. In their seminal paper~\cite{regge}, after appropriate boundary conditions were imposed to the perturbation equations, the stability of Schwarzschild black holes was established. Later, in 1970, Vishveshwara~\cite{vish} discovered a special class of solutions for the perturbation equations of Schwarzschild black holes which are characterized by a peculiar set of boundary conditions: purely outgoing waves at spatial infinity and purely ingoing waves at the event horizon. These solutions, known as quasinormal modes (QNMs), correspond to stable modes which decay in time. Assuming a time dependence of $e^{-i\omega t}$, these modes are characterized by complex frequencies $\omega$ (the quasinormal frequencies) whose imaginary part is negative. In general, only a discrete set of quasinormal frequencies will be allowed. 

Since the discovery of QNMs, several methods have been devised to determine the quasinormal frequencies of a black hole system. The simplest one consists in approximating the black hole potential by a P\"oschl-Teller potential whose analytical solutions are known~\cite{poschl}. Another approximation technique is the WKB method~\cite{will}, which is equivalent to finding the poles of the transmission coefficient of a tunnelling problem in Quantum Mechanics. One can also apply a shooting method to match the asymptotic solutions at some intermediate point, as done by Chandrasekhar and Detweiler~\cite{chandra}.  
Another important method is the direct integration of the perturbation equation in the time domain using light-cone coordinates~\cite{gundlach}. The most accurate method, however, is the continued fraction method developed in 1985 by Leaver~\cite{leaver1,leaver2} and later improved by Nollert~\cite{nollert}. For a detailed review on the available methods, we recommend~\cite{review3} (see also \cite{review1,review2,review4} for other reviews about QNMs).

Leaver's method was inspired by a technique due to Jaffe to calculate the energy eigenvalues of the $H_2^{+}$ ion. It consists in using the Frobenius method of differential equations to write the solution of the perturbation equation as a power series around the event horizon. It can be shown that this series will satisfy the QNM boundary conditions only if a certain equation involving an infinite continued fraction is also satisfied. By solving the continued fraction equation with a root finding algorithm, one is then able to determine the QNMs. Even though Leaver originally applied his method only to massless perturbations of Schwarzschild and Kerr black holes, it can be implemented in a variety of other situations~\cite{review3,review4}.

In this paper, we are particularly interested in QNMs of extremal black holes. Unfortunately, Leaver's original method does not work in such cases, and the reason lies in the fact that the event horizon of an extremal black hole is an irregular singular point of the perturbation equation. Therefore, unlike the case of non-extremal black holes (for which the event horizon is a regular singular point), one does not expect a non-zero radius of convergence of the associated power series around the event horizon. In 1996, however, Onozawa \emph{et al.}~\cite{onozawa} successfully devised a modification of Leaver's original method for an extremal RN black hole which does not rely on expansions around the event horizon. Instead, they expand the solution around an ordinary point of the differential equation. Following the ideas of Onozawa \emph{et al.}, we show in this paper (Secs.~\ref{sec_2} and \ref{cfm}) how to modify the original continued fraction method to calculate the QNMs of an extremal Kerr black hole. We also apply the method to spin-$0$ and spin-$1/2$ charged perturbations around an extremal RN black hole, generalizing the results of Onozawa \emph{et al.} for non-neutral fields. Our numerical results, including a detailed analysis of the so-called damped modes for extremal Kerr black holes and a mode stability investigation, are presented in Sec.~\ref{sec_4}, followed by our final remarks in Sec.~\ref{sec_5}.
   
%================
\section{Field dynamics} \label{sec_2}
%================

According to the black hole uniqueness theorems~\cite{israel, robinson}, under reasonable conditions, the Kerr metric is the only stationary axisymmetric black hole solution of the vacuum Einstein field equations while the RN metric is the only spherically symmetric black hole solution of the Einstein-Maxwell field equations. The first one describes the spacetime of a rotating black hole with mass $M$ and specific angular momentum $a \le M$, and the second one describes an electrically charged black hole of mass $M$ and charge $Q \le M$ (throughout this paper we use units in which $G=c=1$).

The Kerr line element, in Boyer-Lindquist coordinates $(t,r,\theta,\phi)$, is given by
\begin{align}
ds^2 = - \left(1 - \frac{2Mr}{\rho ^2} \right) dt^2 - \frac{4 M r a \sin ^2 \theta}{\rho^2} d\phi \, dt  
+ \frac{\rho^2}{\Delta}dr^2 \nonumber \\ + \rho^2 d \theta^2 + \left( r^2 + a^2 + \frac{2 M r a^2 \sin^2 \theta}{\rho ^2} \right) \sin ^2 \theta \, d \phi ^2, \label{metric}
\end{align}
where $\rho^2 = r^2 + a^2 \cos ^2 \theta$ and $\Delta = r^2 - 2Mr + a^2$. If $a < M$, the roots of the $\Delta$ function, namely $r_+=M+\sqrt{M^2 - a^2}$ and $r_-=M-\sqrt{M^2 - a^2}$, correspond respectively to the event horizon and the Cauchy surface of the black hole. If $a=M$, these two surfaces merge into the surface $r_+=r_-=M$, which corresponds to the event horizon of an extremal Kerr black hole. Finally, if $a < M$, there is no event horizon and the metric corresponds to a naked singularity at $r=0$.

In the early 1970s, Teukolsky derived one of the most notable facts about the Kerr metric: that its perturbations are separable (in the frequency domain) when the Newman-Penrose formalism is employed~\cite{teukolsky}. In fact, let $\psi$ describe any neutral, massless spin-s field ($|s|=0,1/2,1,2$) and write the corresponding wave equation in the Kerr background using the \emph{ansatz} $\psi = R(r)S(\theta) e^{i m \phi - i \omega t}$. As a result, the wave equation reduces to the so-called Teukolsky equations~\cite{teukolsky,teukolsky2} for $S(\theta)$ and $R(r)$: 
\begin{align} 
\frac{1}{\sin \theta} \frac{d}{d \theta} \left( \sin \theta \frac{dS}{d \theta} \right) + \left(a^2 \omega^2 \cos^2 \theta - 2 a \omega s \cos \theta   \right. \nonumber \\ \left. +  s + \lambda - \frac{m^2}{\sin ^2 \theta}  - \frac{2 m s \cos \theta}{\sin ^2 \theta} - s^2 \cot ^2 \theta  \right) S = 0 , \label{teukoang}
\end{align}
and
\be
  \Delta^{-s} \frac{d}{dr} \left(\Delta^{s+1}  \frac{dR }{dr}  \right)  +
\left(
\frac{K^2 -  2is (r-M)K}{\Delta} + X \right)R   =0, \label{teukorad}
\ee 
where $\lambda$ is a separation constant, $K=\omega (r^2 + a^2) -am$, and $ X = 4is \omega r + 2 m a \omega - a^2 \omega^2 - \lambda$. 

The description of a charged black hole system is similar. The RN line element is given by
 \be
\label{RN}
ds^2 = - \frac{\Delta}{r^2} dt^2 + \frac{r^2}{\Delta} dr^2 + r^2 \left( d \theta ^2 + \sin ^2 \theta  \, d \phi ^2 \right),
\ee
where $\Delta = r^2 - 2Mr + Q^2$. Its event horizon is located at $r_+=M+\sqrt{M^2 - Q^2}$, which reduces to $r_+=M$ in the extremal case ($Q=M$). Furthermore, the study of spin-$0$ and spin-$1/2$ electrically charged, massless fields around a RN black hole can be simplified if the same \emph{ansatz} used for Kerr perturbations is employed. Indeed, by doing so, one can show that the Klein-Gordon and the Dirac equations, in the RN metric, can be separated and recast as new equations which are analogous to the Teukolsky equations~\cite{jing,mauricio_alberto}. The angular equation assumes the form \eqref{teukoang} with $a\omega = 0$, while the radial one assumes the form \eqref{teukorad} with $K = \omega r^2 -qQr$ and $X = 4is \omega r - 2isqQ - \lambda$ ($q$ denotes the charge of the field).

In order to solve the perturbation equations~\eqref{teukoang} and \eqref{teukorad}, one needs appropriate boundary conditions. For the angular function, it is natural to require regularity at the singular points $\theta = 0$ and $\theta = \pi$, transforming eq.~\eqref{teukoang} into a Sturm--Liouville problem for the separation constant. For a given set of parameters $a\omega$, $s$ and $m$, only a discrete set (indexed by the new parameter $\ell$) of constants $\lambda$ is allowed. The associated functions are called the spin-weighted spheroidal harmonics~\cite{teukolsky2,spinwei}.
In the particular case of $a\omega = 0$, valid for the RN metric, these angular functions reduce to the spin-weighted spherical harmonics, for which the separation constant can be determined analytically as $\lambda = (\ell - s)(\ell + s + 1)$. Additionally, the parameters satisfy the following constraints: $m$ is an integer (bosonic case) or a half-integer (fermionic case), $\ell$ is a non-negative integer (bosonic case) or a non-negative half-integer (fermionic case), $\ell \ge |s|$, and $-\ell \le m \le \ell$.

 Regarding the radial equation, on the other hand, we impose the usual boundary conditions for QNMs: purely ingoing modes at the event horizon (since no classical perturbation can escape the black hole) and purely outgoing modes in the asymptotic limit $r \rightarrow \infty$. In these limits, eq.~\eqref{teukorad} can be solved analytically, yielding the following boundary conditions (for an extremal black hole): 
\begin{align}
&R(r) \propto e^{\frac{J_0}{r-M}}(r-M)^{J_1}, \quad &\text {when }r \rightarrow M, \label{bc1} \\
&R(r) \propto r^{J_1 + J_2}e^{+i\omega r}, \quad &\text {when }r \rightarrow \infty, \label{bc2}
\end{align}
where $J_0=iM(2M \omega - m)$, $J_1=-2s - 2iM\omega$, and $J_2=-1 + 4 i M \omega$ in the case of an extremal Kerr black hole, and $J_0=iM( M\omega - qQ)$, $J_1=-2s - 2iM\omega + iqQ$, and $J_2=-1 + 4 i M \omega - 2iqQ$ for an extremal RN black hole.

%================
\section{Continued Fraction method} \label{cfm}
%================

As explained in the introduction, the original continued fraction method devised by Leaver, which consists in expressing the solution of the radial Teukolsky equation as a power series around the event horizon, works only for non-extremal black holes. Therefore, inspired by the ideas of Ref.~\cite{onozawa} for neutral perturbations around an extremal RN black hole, we expand the solutions of the radial equation \eqref{teukorad} around the ordinary point $r=2M$. More precisely, we write
\begin{align} \label{exp2}
R(r)= e^{i \omega r} e^{\frac{J_0}{r-M}}(r-M)^{J_1} r^{J_2} \sum_{n=0}^{\infty}a_n\left(\frac{r-2M}{r} \right)^n,
\end{align}
where the expansion coefficients $a_n$ satisfy a five-term recurrence relation (the last equation holds for $n \ge 3$):
\begin{align} 
&\alpha_1 a_2 + \beta_1 a_1 + \gamma_1 a_0 = 0, \label{rec0}  \\
&\alpha_2 a_3 + \beta_2 a_2 + \gamma_2 a_1 + \delta_2 a_0 = 0, \label{rec1}  \\
&\alpha_n a_{n+1} + \beta_n a_n + \gamma_n a_{n-1} + \delta_n a_{n-2} +\epsilon_n a_{n-3}= 0. \label{5term}
\end{align}
The coefficients $\alpha_n$, $\beta_n$, $\gamma_n$, $\delta_n$ and $\epsilon_n$, for both Kerr and RN extremal black holes, are given explicitly in the Appendix.
 
 Note that the boundary conditions \eqref{bc1} and \eqref{bc2} are automatically satisfied provided that the power series is everywhere convergent. One possible way to test convergence is to analyse the large-$n$ behaviour of $a_{n+1}/a_n$. Using~\eqref{5term}, it is possible to find four different asymptotic solutions, namely
\be
\frac{a_{n+1}}{a_n} = 1 \pm 2\sqrt{\frac{-iM\omega}{n}} + \mathcal{O}\left(n^{-1}\right) \label{nol1},
\ee
and
\be
\frac{a_{n+1}}{a_n} =-1 \pm 2\sqrt{\frac{-J_0}{M n}}  + \mathcal{O}\left(n^{-1}\right).  \label{nol2}
\ee
Since $|a_{n+1}/a_n| \rightarrow 1$ in all four cases, the ratio test guarantees that the sum in \eqref{exp2} is convergent as long as $M < r < \infty$. However, convergence at $r=M$ and at $r=\infty$, i.e.~absolute convergence of the sum $\sum a_n$, is also needed to ensure the fulfilment of the QNM boundary conditions. With the help of Raabe's test, we find that the solutions associated with the minus sign in \eqref{nol1} or the plus sign in \eqref{nol2} are compatible with QNMs. The only exception happens when $iM\omega$ or $J_0$ are positive real numbers, since then the coefficient of the $n^{-1/2}$ term becomes purely imaginary~\cite{nollert}.

Equation \eqref{5term}, being a five-term linear recurrence relation, possesses four independent solutions, each one yielding one of the four possible asymptotic behaviours discussed above. (In other words, one has the freedom to choose $a_0$, $a_1$, $a_2$ and $a_3$, from which all the other $a_n$'s will be determined.) Futhermore, equations \eqref{rec0} and \eqref{rec1} can be thought as boundary conditions for \eqref{5term}, reducing the degrees of freedom from 4 to 2. Equivalently, one can perform two successive Gaussian eliminations in \eqref{5term} to transform eqs.~\eqref{rec0}--\eqref{5term} into a three-term recurrence relation for $a_n$: 
\be \label{3term}
\alpha_n '' a_{n+1} + \beta_n '' a_n + \gamma_n '' a_{n-1} = 0, \qquad n \ge 1,
\ee
where the new recurrence coefficients $\alpha_n ''$, $\beta_n ''$ and $\gamma_n ''$ are obtained recursively from the original ones [see eqs.~\eqref{gauss1} and \eqref{gauss2} in the Appendix].

For a given frequency $\omega$, this three-term recurrence relation possesses two linearly independent solutions. In general, these solutions will produce divergent sums $\sum a_n$ and, therefore, will be incompatible with the QNM boundary conditions. 
Consequently, in order to determine the QNMs, we need to find the frequencies $\omega$ for which one of the solutions of \eqref{3term} will satisfy the convergence condition discussed above (and related to the asymptotic behaviour of $a_{n+1}/a_n$). 

Note that the absolute convergence of the series $\sum a_n$ is equivalent to the convergence of the sums $\sum c_n := \sum a_{2n} $ and $\sum d_n :=  \sum a_{2n+1}$. By rearranging and manipulating the three-term recurrence relation \eqref{3term} for three successive integers, one can eliminate all even-numbered  (or, similarly, all odd-numbered) terms from the equations to decouple $c_n$ from $d_n$, obtaining
\begin{align}
 \alpha_n ^ e c_{n+1} +  \beta_n ^e c_n +  \gamma_n ^e c_{n-1} = 0, \qquad n \ge 1, \label{cn}  \\
 \alpha_n ^o d_{n+1} + \beta_n ^o d_n + \gamma_n ^o d_{n-1} = 0, \qquad n \ge 1, \label{dn} 
\end{align}
where the new recurrence coefficients are given in terms of the original ones in eqs.~\eqref{odd} and \eqref{even}. See the Appendix for all the details of the decoupling process. We remark that the decoupling in Ref.~\cite{onozawa} assumes that the coefficient $\delta_n$ is zero, while our decoupling works for $\delta_n \neq 0$.

We now invoke the theory of three-term recurrence relations, and its relation to continued fractions~\cite{gaut}, to determine the quasinormal frequencies. More precisely, we use the fact that $a_n$ is a minimal solution of \eqref{3term} if, and only if, the infinite continued fraction $ \frac{\gamma_1''}{\beta_1'' -}\frac{\alpha_1''\gamma_2''}{\beta_2'' -}\frac{\alpha_2''\gamma_3''}{\beta_3'' -} \dots$ converges. Furthermore, in case of convergence, the minimal solution satisfies
\be  
\frac{a_1}{a_0}=  - \frac{\gamma_1''}{\beta_1'' -}\frac{\alpha_1''\gamma_2''}{\beta_2'' -}\frac{\alpha_2''\gamma_3''}{\beta_3'' -} \dots \label{cf1}
\ee
Of course, similar expressions will hold for $c_1/c_0$ and $d_1/d_0$ provided that $c_n$ and $d_n$ are, respectively, minimal solutions of the recurrence relations \eqref{cn} and \eqref{dn}. As noted by Leaver for non-extremal black holes, minimal solutions are exactly the ones which correspond to the QNM boundary conditions. 

 These continued fractions are not all independent; they can be related through~\eqref{rec0}. More precisely, after dividing \eqref{rec0} by $a_0$, we use relation \eqref{cf1} above (and the analogous expression for $c_1/c_0$) to obtain
\begin{align} 
\alpha_1 \frac{c_1}{c_0} + \beta_1 \frac{a_1}{a_0} + \gamma_1 = - \alpha_1 \frac{\gamma_1^o}{\beta_1^o -}\frac{\alpha_1^o\gamma_2^o}{\beta_2^o -}\frac{\alpha_2^o\gamma_3^o}{\beta_3^o -} \dots \nonumber \\ - \beta_1 \frac{\gamma_1''}{\beta_1'' -}\frac{\alpha_1''\gamma_2''}{\beta_2'' -}\frac{\alpha_2''\gamma_3''}{\beta_3'' -} \dots  + \gamma_1 = 0. \label{cond1}  
\end{align}
Alternatively, we could have used \eqref{rec0} and \eqref{rec1} to obtain 
\be \label{cond2}
\frac{\alpha_2}{\beta_1} \frac{d_1}{d_0} + \frac{\gamma_2}{\beta_1} - \frac{\delta_2 + \beta_2 \frac{c_1}{c_0}}{\gamma_1 + \alpha_1 \frac{c_1}{c_0}} = 0,
\ee
which relates the even and odd numbered sequences after we substitute $c_1/c_0$ and $d_1/d_0$ by their associated continued fractions. Both equations~\eqref{cond1} and~\eqref{cond2} are equivalent and, therefore, the results of any of them can be used as a consistency check for the results obtained with the other one.  

For an extremal RN black hole, since the separation constant $\lambda$ can be determined analytically, one can directly solve the continued fraction equation \eqref{cond1} to determine the quasinormal modes. However, for an extremal Kerr black hole, since not only $\omega$, but also the separation constant $\lambda$ is not known \emph{a priori}, we also need to solve the angular Teukolsky equation numerically. The mathematical treatment is exactly the same one used for non-extremal black holes~\cite{leaver1}: we define $u=\cos \theta$ and expand the solution $S(u)$ around the regular singular point $u=-1$ using the \emph{ansatz}
 \begin{align} \label{ang_exp}
S(u)=e^{a\omega u}(1-u)^{\frac{|m+s|}{2}}\sum_{n=0}^{\infty}b_n\left(1+ u \right)^{n + \frac{|m-s|}{2} },
\end{align}
where the expansion coefficients $b_n$ satisfy the following three-term recurrence relation:
\begin{align} 
&\alpha_0^{\theta} b_1 + \beta_0 ^{\theta}b_0 = 0, \label{ang_rec0}  \\
&\alpha_n ^{\theta} b_{n+1}+ \beta_n ^{\theta} b_n + \gamma_n ^{\theta} b_{n-1}= 0,  \qquad n \ge 1, \label{ang_rec_3term} 
\end{align}
with $\alpha_n^{\theta}$, $\beta_n^{\theta}$ and $\gamma_n^{\theta}$ given by 
\begin{align}
\alpha_n ^{\theta} &= -4(1+n)(1+n+|m-s|); \\
\beta_n ^{\theta} &= -2 a^2 \omega ^2+ \left| m-s \right|  (-4 a \omega +\left| m+s\right| +2 n+1) \nonumber \\
&-2 (2 a \omega +\lambda )-4 a \omega  (2 n+s)+(2 n+1) \left| m+s\right| \nonumber \\
& +m^2+2 n (n+1)-s (s+2); \\
\gamma_n ^{\theta} &= 2a\omega \left[2(n+s) + |m-s| + |m+s|\right].
\end{align}
By invoking the fact that the sum in \eqref{ang_exp} is convergent only if $b_n$ is a minimal solution of \eqref{ang_rec_3term}, we find
\be \label{condang} 
\frac{b_1}{b_0} = \frac{\beta_0 ^{\theta}}{\alpha_0^{\theta}} = - \frac{\gamma_1^{\theta}}{\beta_1^{\theta} -}\frac{\alpha_1^{\theta}\gamma_2^{\theta}}{\beta_2^{\theta} -}\frac{\alpha_2^{\theta}\gamma_3^{\theta}}{\beta_3^{\theta} -} \dots,
\ee
which, together with \eqref{cond1}, can be used to determine the QNMs of an extremal Kerr black hole.

%================
\section{Numerical Results} \label{sec_4}
%================

We now proceed to employ the technique described in the previous section to determine the QNMs of extremal black holes. For a RN black hole, we directly solve eq.~\eqref{cond1} for the unknown $\omega$. For a Kerr black hole, on the other hand, equations \eqref{cond1} and \eqref{condang} form a system of coupled, algebraic equations for $\lambda$ and $\omega$ and, therefore, need to be solved simultaneously. In both cases, we truncate the continued fractions at some sufficiently large order and use a root-finding algorithm to solve the corresponding equation(s) for the unknown(s). We then increase the number of terms in the continued fractions and repeat the procedure until the desired accuracy is attained. We also compare our results for extremal black holes with the quasinormal frequencies of near extremal black holes ($a/M=0.999$ or $Q/M=0.999$), which we calculate using Leaver's original method. Our implementation of Leaver's method follows the steps discussed in~\cite{leaver1} for a Kerr black hole and in~\cite{davi} for a RN black hole, and, therefore, is not repeated here. In order to test it, we have used the tabulated values of Refs.~\cite{leaver1,cardoso_data,davi}, finding excellent agreement between them and our implementation.   

As explained in Refs.~\cite{leaver1,review3, spinwei}, the spheroidal harmonics are invariant under the transformation ($s \rightarrow -s$, $\lambda \rightarrow \lambda + 2s$). This means that the Kerr black hole system is symmetric with respect to the transformation ($m \rightarrow - m$, $\omega \rightarrow -\omega^*$, $\lambda \rightarrow \lambda^*$). The RN black hole system, on the other hand, is symmetric with respect to the transformation ($q \rightarrow - q$, $\omega \rightarrow -\omega^*$). Consequently, without loss of generality, we can assume (unless otherwise stated) that $s$ is non-positive and that $Re(M\omega)>0$.   

\begin{table*}[t]
\centering \caption{Comparison of the QNM frequencies $M\omega$ and the associated separation constants $\lambda$ between a near extremal ($a/M=0.999$) and an extremal Kerr black hole for $m=0$. For each set of parameters $s$ and $\ell$, we have calculated the least damped mode ($n=0$) and the first overtone ($n=1$).}  \vskip 2pt
\begin{tabular}{@{}c|c|c|cccc|cccc@{}}
\hline 
$s$&$\ell$&$n$ &\multicolumn{4}{c|}{$a/M=0.999$}  & \multicolumn{4}{c}{$a/M=1$}\\ \hline
  & &   & $\text{Re}(M\omega)$     &$\text{Im}(M\omega)$ & $\text{Re}(\lambda)$ &$\text{Im}(\lambda)$ & $\text{Re}(M\omega)$ &$\text{Im}(M\omega)$ & $\text{Re}(\lambda)$ &$\text{Im}(\lambda)$  \\
 $0$ & $0$ & $0$   & $0.110265$ & $- 0.089439$  &  $ -0.001378$ & $ 0.006564$  & $ 0.110245$ & $ - 0.089433 $ & $ -0.001380$ & $ 0.006575 $  \\
 $0$ & $0$ & $1$ & $ 0.062498$ & $-0.318852 $  &  $ 0.032404$ & $0.013144$  & $ 0.062473 $ & $ - 0.318840 $ & $ 0.032468$ & $ 0.013164 $ \\
 $0$ & $1$ & $0$    & $0.314946$ & $- 0.081771 $  &  $ 1.944568$ & $ 0.030907$  & $ 0.314986$ & $ -  0.081714 $ & $ 1.944436$ & $ 0.030952 $  \\
 $0$ & $1$ & $1$ & $ 0.281435$ & $-0.253809 $  &  $ 1.991284$ & $ 0.085574$  & $0.281392$ & $ - 0.253686 $ & $ 1.991243$ & $0.085691 $  \\           
 $-1$ & $1$ & $0$    & $0.274777$ & $- 0.075305$  &  $ 1.972061$ & $ 0.016634$  & $ 0.274828$ & $ - 0.075232 $ & $ 1.971989 $ & $ 0.016655 $  \\
  $-1$ & $1$ & $1$         & $  0.240106$ & $- 0.234599 $  &  $ 1.999206$ & $ 0.044984$  & $0.240053 $ & $ - 0.234445 $ & $ 1.999185 $ & $ 0.045035 $  \\           
  $-1$ & $2$ & $0$    & $0.500902$ & $- 0.079434$  &  $ 5.907613 $ & $ 0.029813$  & $ 0.501013 $ & $ - 0.079365$ & $ 5.907384 $ & $ 0.029852 $  \\
  $-1$ & $2$ & $1$         & $ 0.479397$ & $-  0.241175 $  &  $  5.934462$ & $ 0.087017$  & $0.479437 $ & $ - 0.240991 $ & $ 5.934283 $ & $ 0.087130 $  \\           
  $-2$ & $2$ & $0$    & $0.424998 $ & $- 0.071899 $  &  $3.907902$ & $ 0.032246$  & $0.425145$ & $ - 0.071806 $ & $ 3.907644 $ & $ 0.032281 $  \\
  $-2$ & $2$ & $1$ & $ 0.402689$ & $-0.218528 $  &  $ 3.940442$ & $ 0.092568$  & $ 0.402744 $ & $ - 0.218283 $ & $ 3.940242$ & $  0.092663 $  \\           
 $-2$ & $3$ & $0$    & $0.664945 $ & $- 0.076814 $  &  $9.856062$ & $ 0.033400$  & $ 0.665132 $ & $ - 0.076735 $ & $ 9.855691 $ & $ 0.033441 $  \\
 $-2$ & $3$ & $1$ & $ 0.649315$ & $-  0.231889 $  &  $ 9.877945$ & $0.098717$  & $ 0.649443$ & $ - 0.231668 $ & $ 9.877614 $ & $ 0.098836 $  \\           
\hline 
\end{tabular}
\label{table1}
\end{table*}  

\subsection {Extremal Kerr black holes}

We start our analysis with $m=0$ perturbations around a Kerr black hole. The quasinormal frequencies of an extremal black hole, together with the quasinormal frequencies of a near extremal black hole, are presented in table \ref{table1}. The associated separation constants $\lambda$ are also shown. Note that the relative difference between the results is minimal: for $Re(M\omega)$, it ranges from $0.008$\% ($s=-1$, $\ell=2$, $n=1$) to  $0.04$\%  ($s=\ell=0$, $n=1$),  and for $Im(M\omega)$ it ranges from $0.004$\% ($s=\ell=0$, $n=1$) to  $0.13$\%  ($s=-2$, $\ell=2$, $n=0$). On the one hand, our results show that the use of Leaver's method for near extremal Kerr black holes is an excellent way to estimate the actual values for the extremal case. On the other hand, our method provides the most accurate technique available for the determination of QNMs of extremal Kerr black holes.

\begin{figure}
\begin{center}
\includegraphics[width=8.6cm]{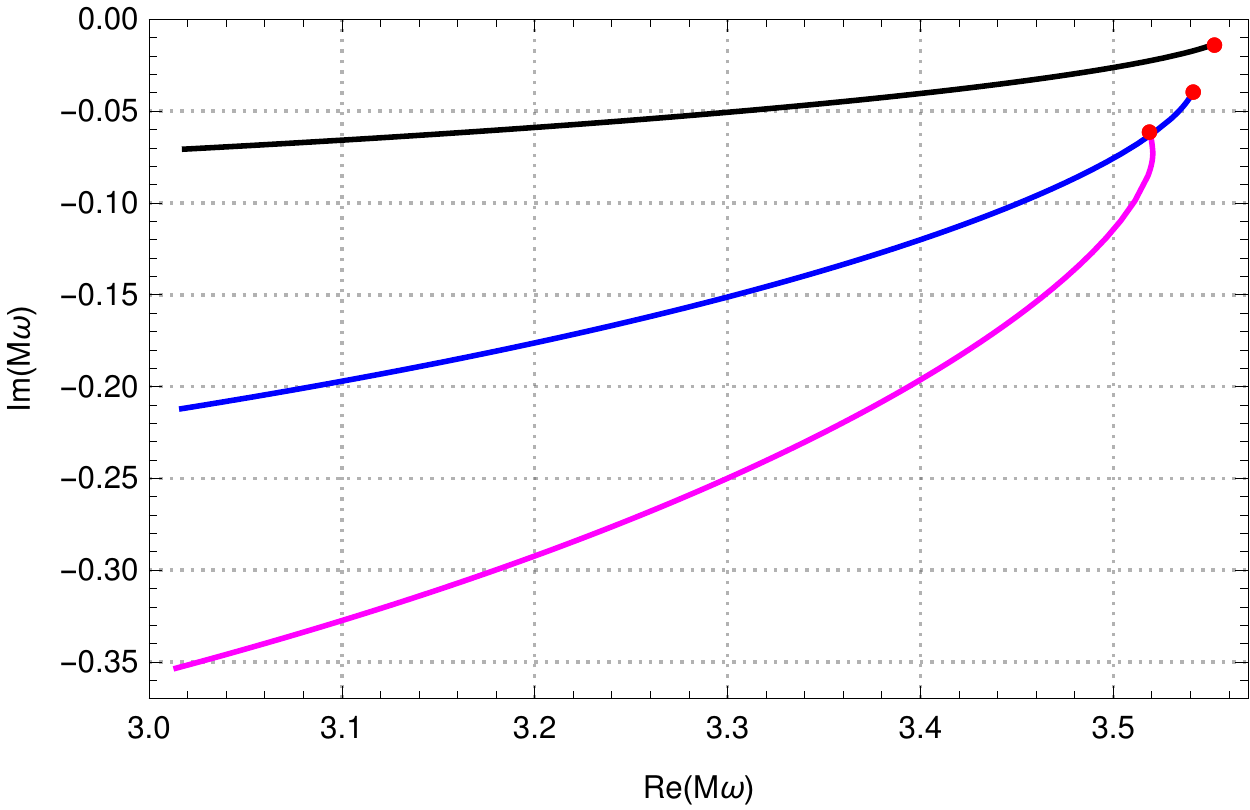}\\
\includegraphics[width=8.6cm]{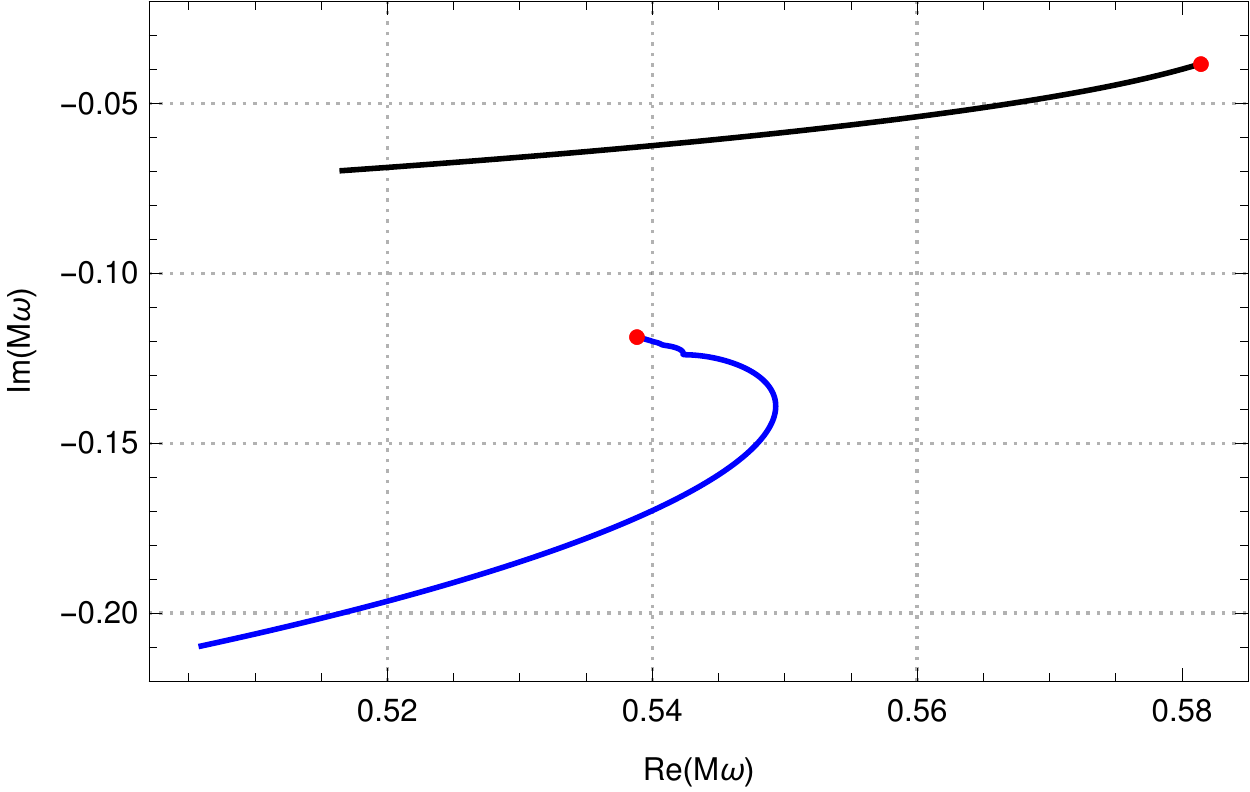}
\caption{(Colors online.) Parametric plots of the real and imaginary parts of the first damped modes for ($s=-2$, $\ell=2$, $m=1$) (upper plots) and ($s=0$, $\ell=10$, $m=7$) (lower plots).  All lines start at $a/M=0.9$, end at $a/M=0.99999$, and are obtained using Leaver's original method; red marks, on the other hand, correspond to extremal black holes and are obtained using the method discussed in this paper. 
\label{fig1}
}
\end{center}
\end{figure}

\begin{figure}
\begin{center}
\includegraphics[width=8.6cm]{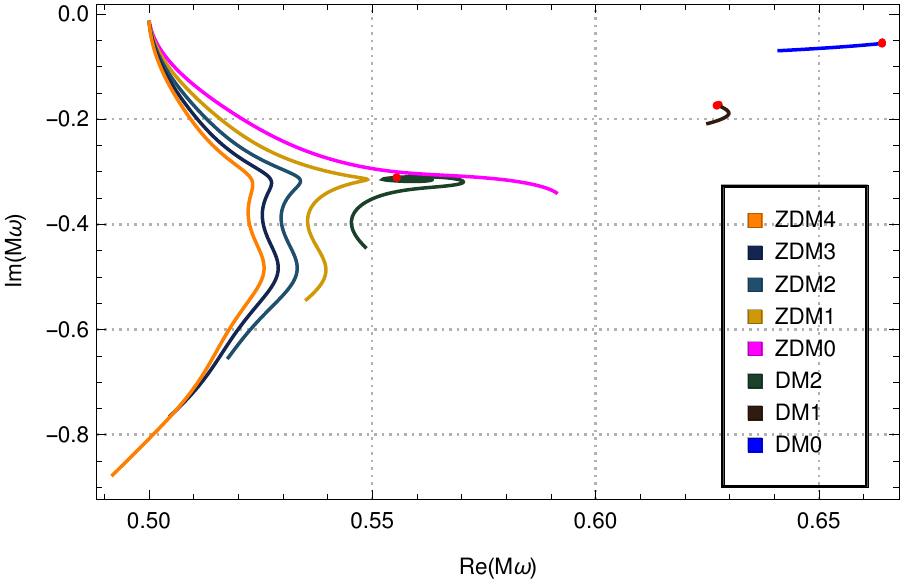}\\
\includegraphics[width=8.6cm]{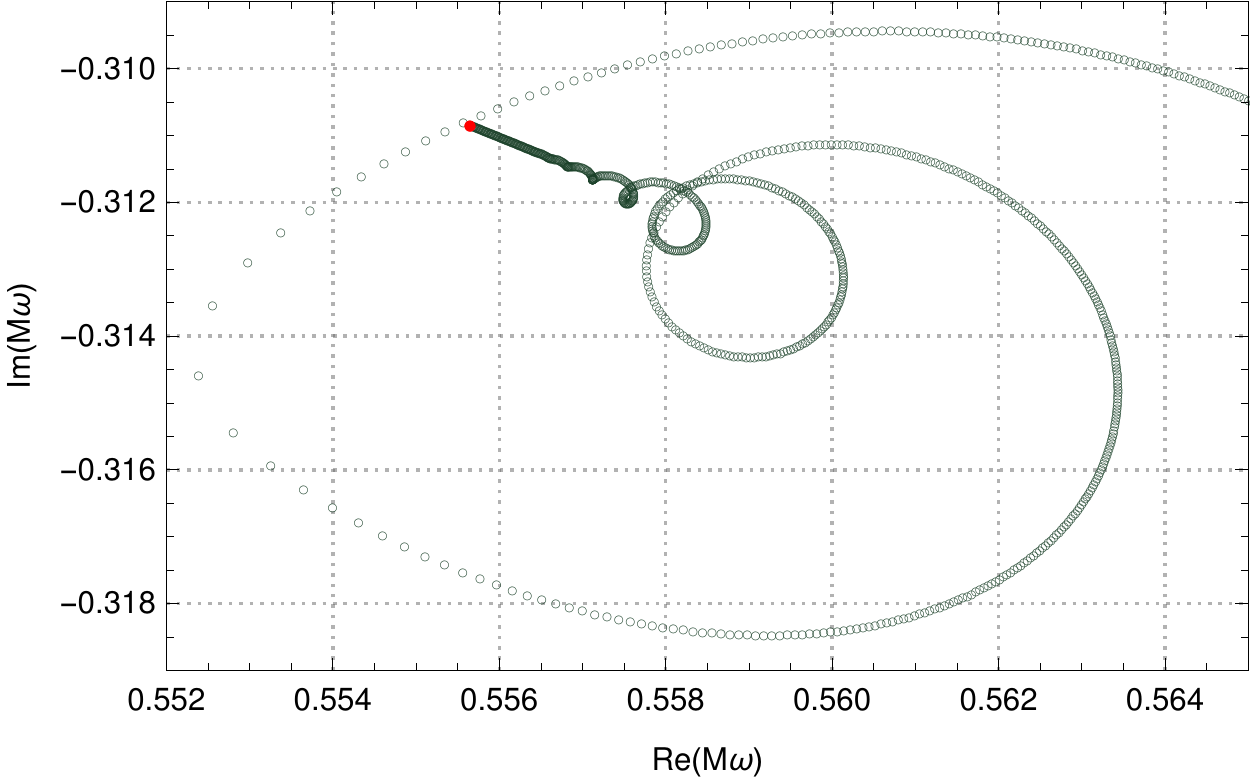}
\caption{(Colors online.) Upper plot is a parametric plot of the real and imaginary parts of the first damped and zero-damping modes for ($s=0$, $\ell=2$, $m=1$). All lines start at $a/M=0.95$, end at $a/M=0.99999$, and are obtained using Leaver's original method. Note that, as extremality is approached, the zero-damping modes converge towards the purely real value $m/2$, while the damped modes approach complex numbers with non-zero imaginary part. The damped modes at extremality, indicated by red dots, are obtained with the method described in this paper. The bottom plot is a close-up view of the third damped mode (and its spiralling behaviour). 
\label{fig2}
}
\end{center}
\end{figure} 

\begin{figure*}
\begin{center}
\begin{tabular}{cc}
\includegraphics[width=8.6cm]{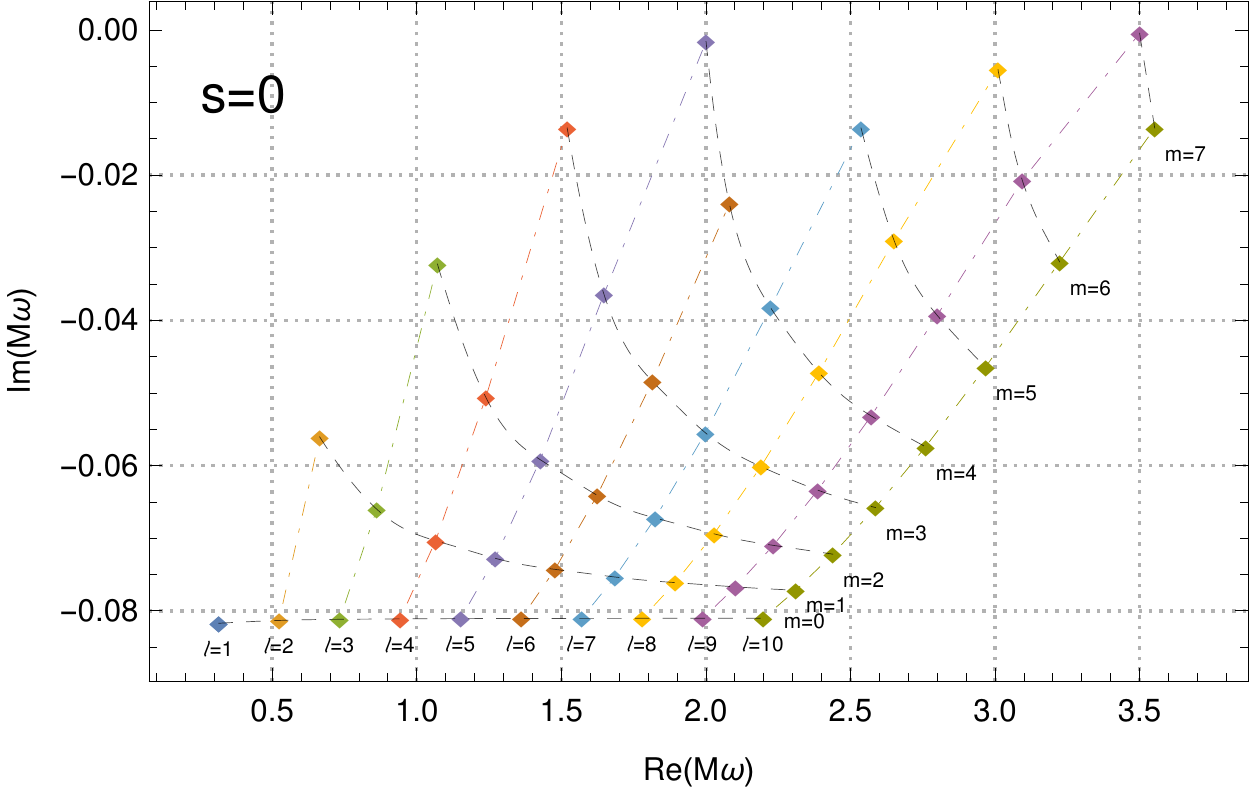}&
\includegraphics[width=8.6cm]{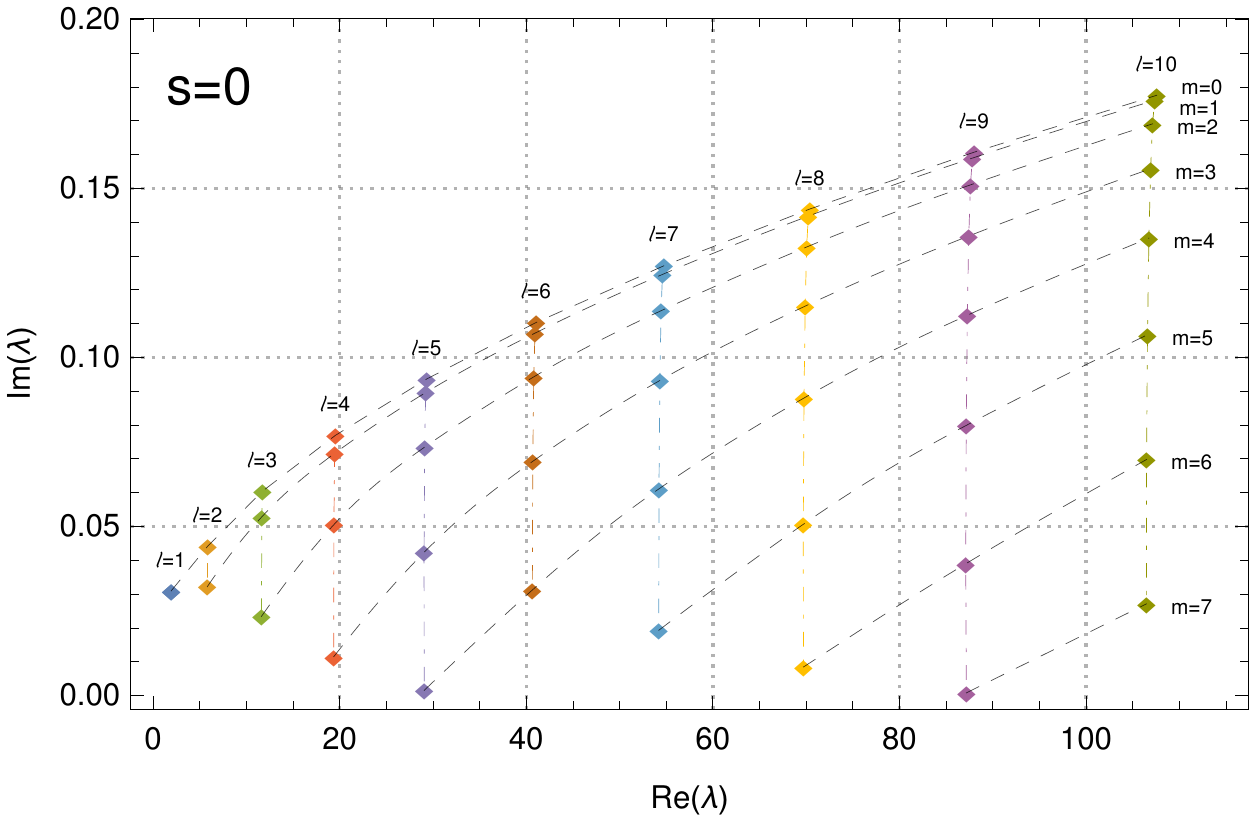} \\
\includegraphics[width=8.6cm]{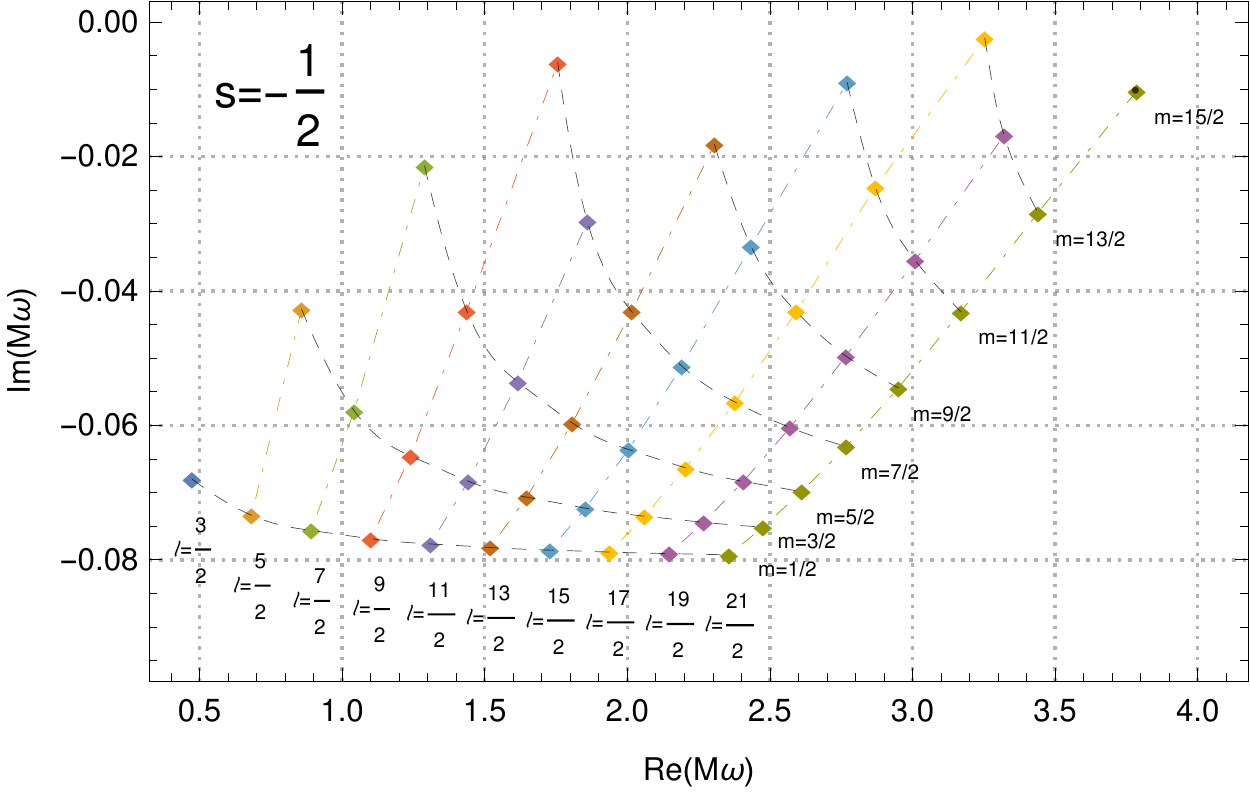}&
\includegraphics[width=8.6cm]{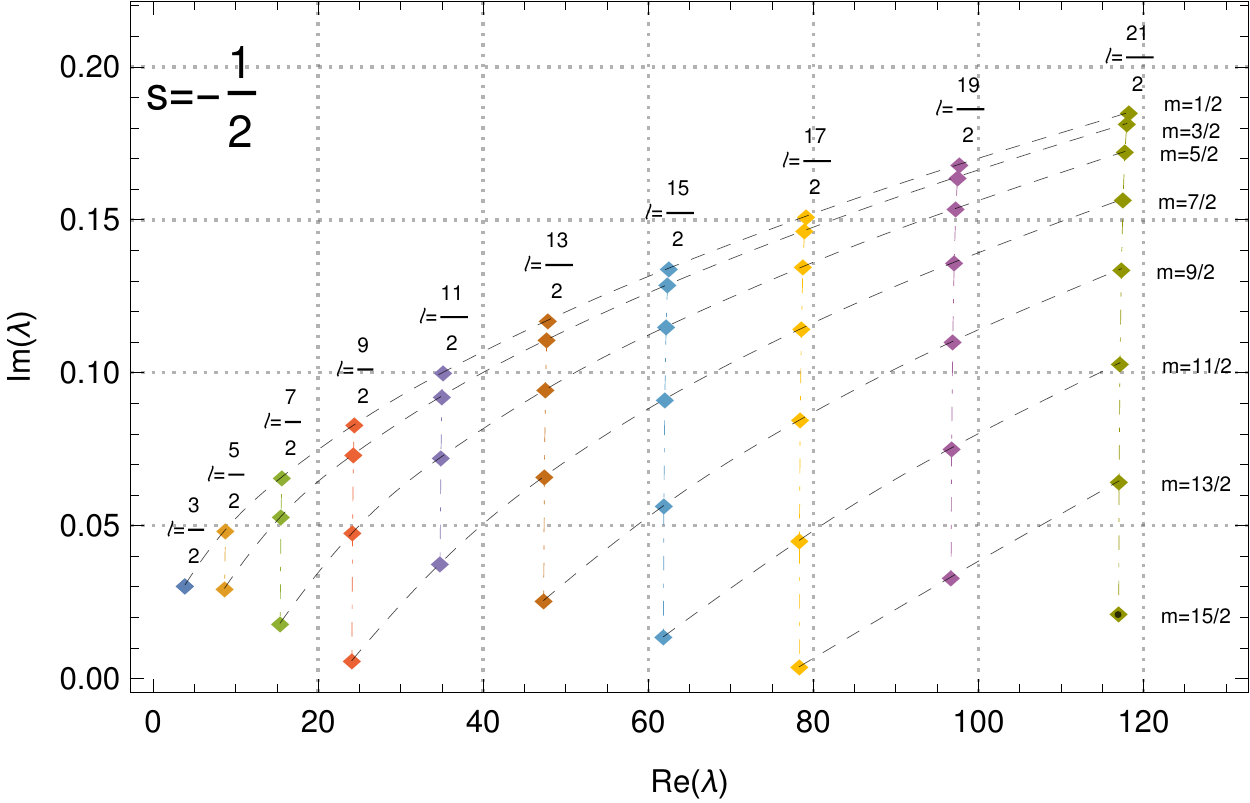} \\
\includegraphics[width=8.6cm]{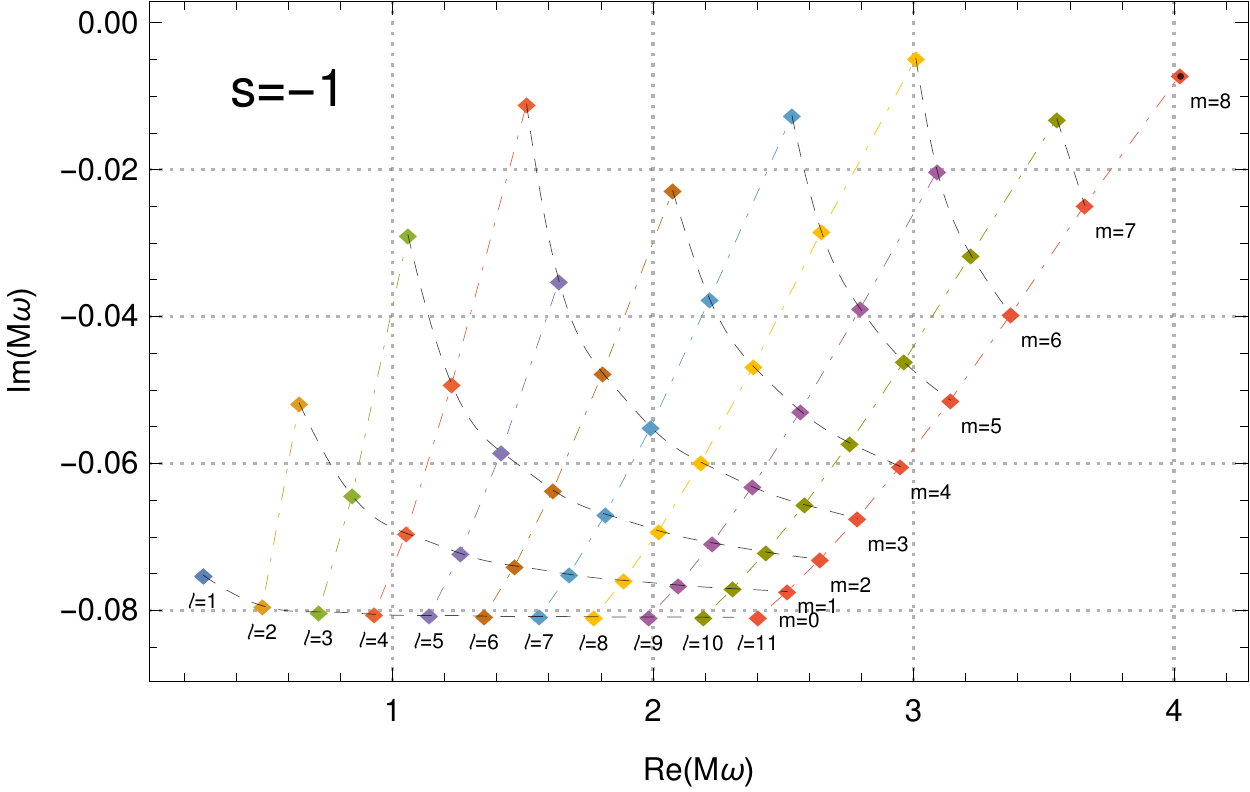}&
\includegraphics[width=8.6cm]{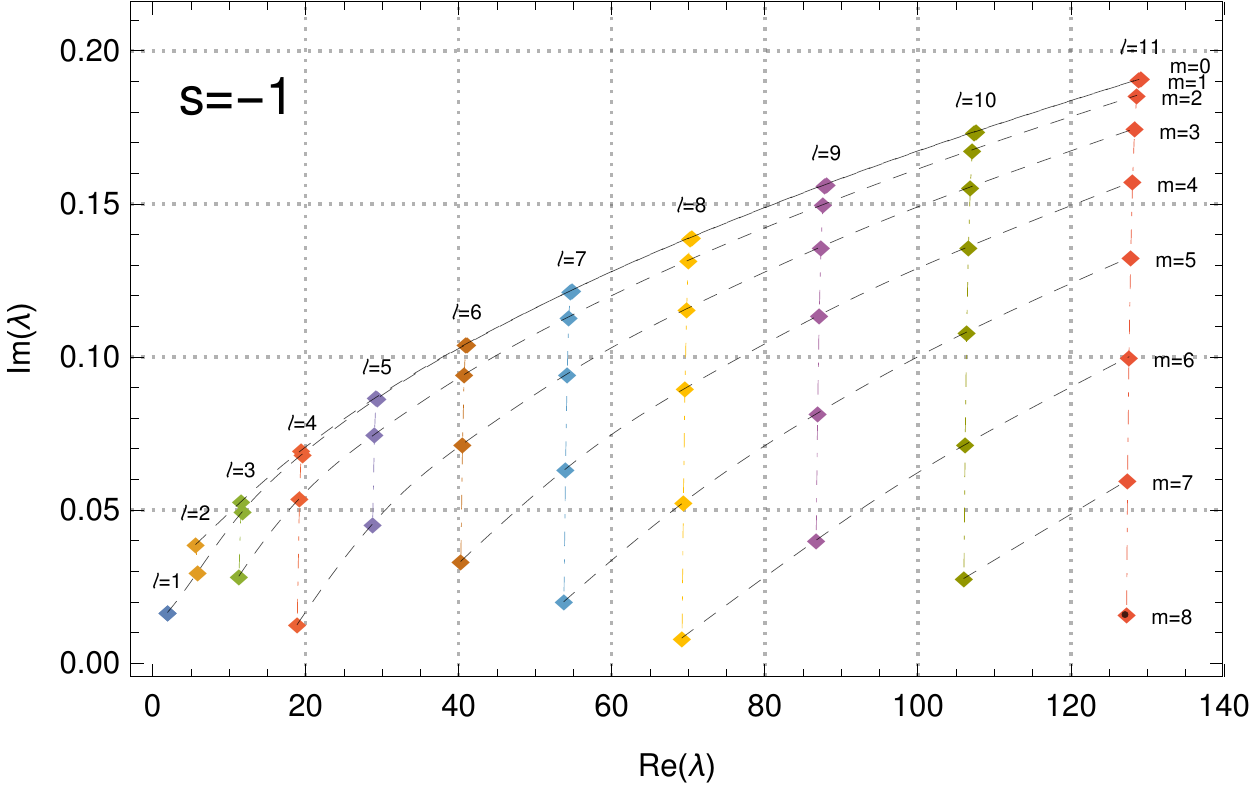} \\
\includegraphics[width=8.6cm]{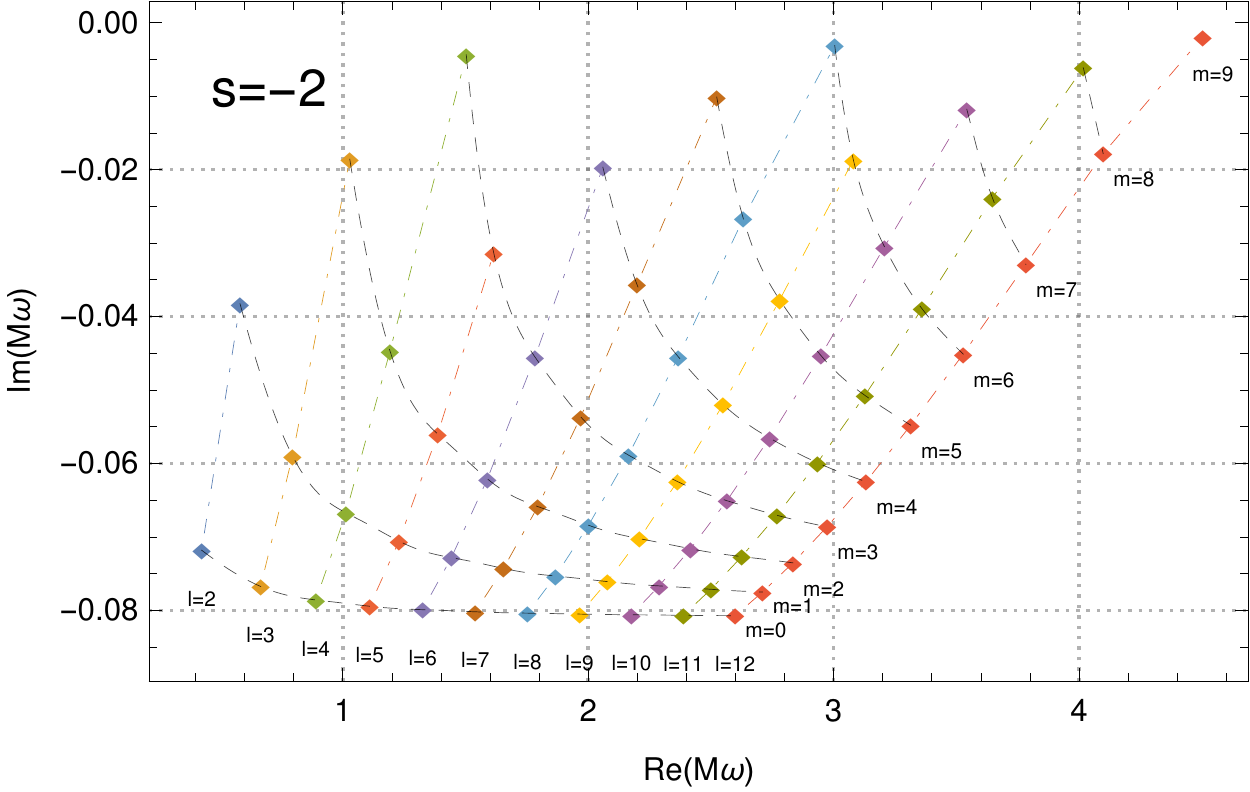}&
\includegraphics[width=8.6cm]{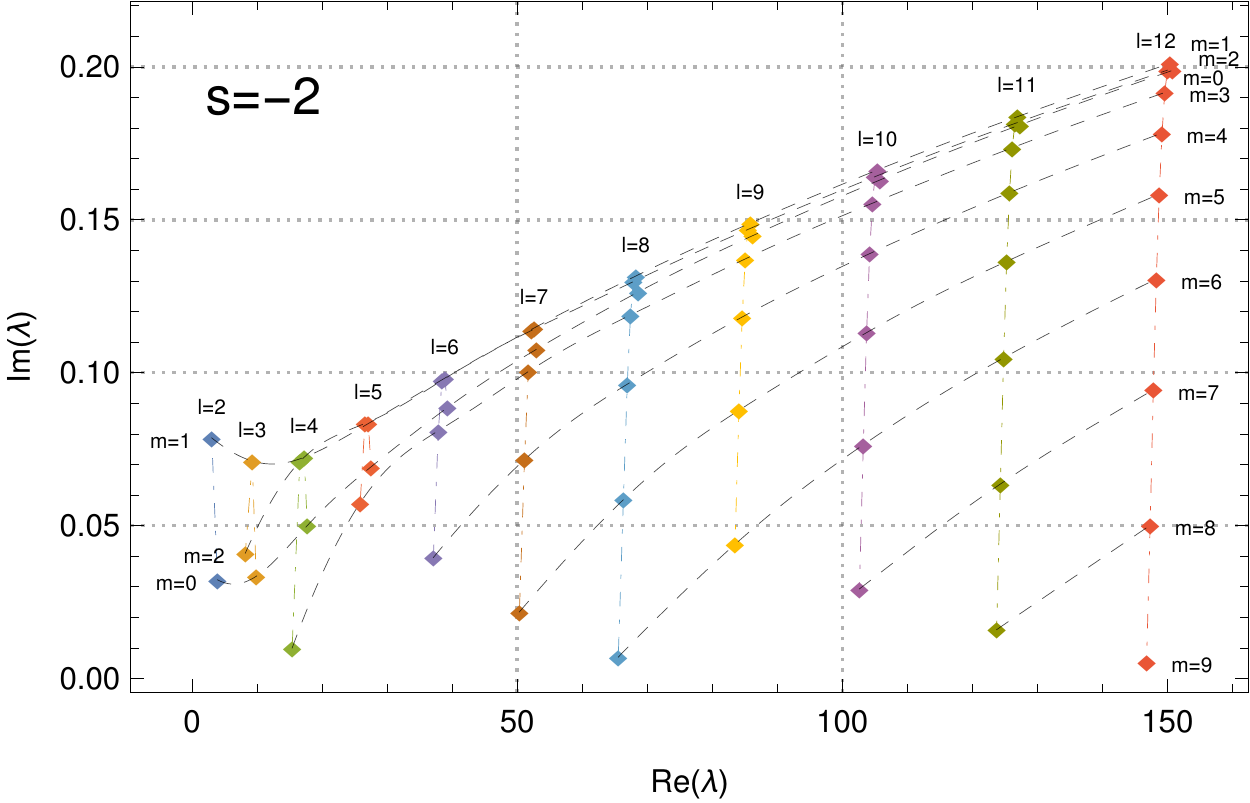}
\end{tabular}
\caption{(Colors online.) Real and imaginary parts (left plots) of the frequencies of the most stable damped modes, and the associated separation constants (right plots), for scalar, Dirac, electromagnetic and gravitational perturbations. Note the regular behaviour of the QNMs as the parameters $\ell$ and $m$ are varied.
\label{fig3}
}
\end{center}
\end{figure*}

We now turn our attention to Kerr QNMs with $m>0$ (corotating modes). As first predicted by Detweiler in~\cite{detweiler} after an analytical treatment of the near extremal regime, when the black hole approaches extremality, its quasinormal frequencies approach the purely real value $M\omega = m /2$. This behaviour, further studied in Refs.~\cite{ferrari,glampe,cardosox,hod00,branching1,branching2}, suggests the existence of QNMs with exactly vanishing imaginary parts for extremal Kerr black holes. Ideally, we would like to apply the continued fraction method to determine if these QNMs modes with vanishing imaginary parts are indeed part of the quasinormal spectrum of extremal black holes. However, when $M\omega = m/2$, there is no guarantee that the method will work, since we have $J_0 = 0$ and, therefore, \eqref{bc1} is not correct anymore. Nevertheless, as we increase the number of terms in the continued fractions, the solutions of eqs.~\eqref{cond1} and \eqref{condang} seem to indeed approach $M\omega = m/2$.

     Besides these zero-damping modes, near extremal Kerr black holes exhibit, for some set of parameters $\ell$ and $m$, QNMs whose imaginary part approaches a non-zero value as $a/M \rightarrow 1$~\cite{branching1,hod01,branching2}. These modes, referred to as damped modes, can be obtained for extremal black holes using the method discussed in this paper. In particular, the damped modes we obtain for $s=-2$, $\ell=2$, $m=1$ and $s=0$, $\ell=10$, $m=7$, presented in Fig.~\ref{fig1}, are compatible with those obtained in Figs.~8 and 9 of Ref.~\cite{branching2} for near extremal black holes.

\begin{figure}
\begin{center}
\includegraphics[width=8.6cm]{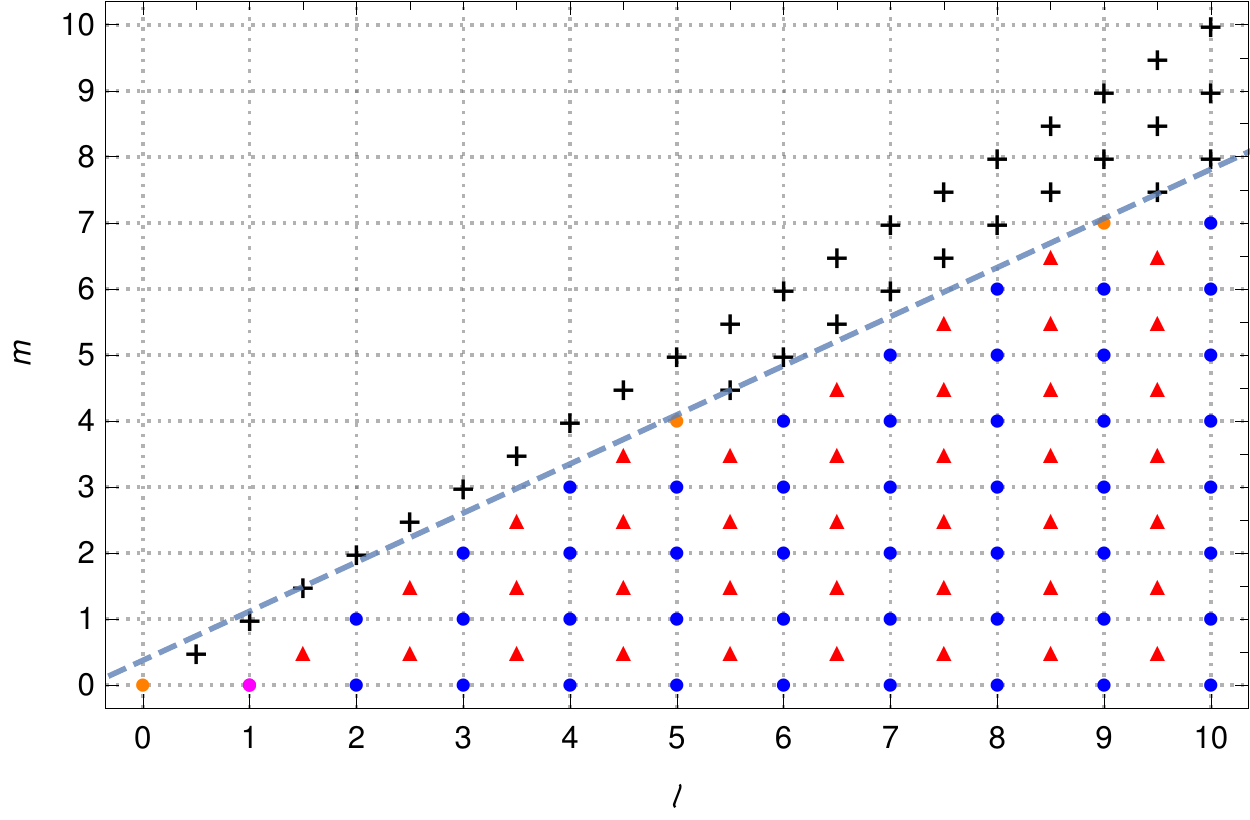}
\caption{(Colors online.) Existence of DMs is summarized in this plot. Pairs $(\ell,m)$ for which no DMs are allowed are represented by a black cross. Red triangles indicate the existence of DMs for $s=-1/2$. Blue dots indicate the existence of DMs for all bosons ($s=0$, $1$ and $2$ simultaneously), magenta dots indicate the existence of DMs for $s=0$ and $1$, but no DMs for $s=2$, and, finally, orange dots indicate the existence of DMs only for $s=0$. The dashed line corresponds to the theoretical result $m/(\ell+1/2) = 0.744$ obtained using the eikonal approximation. 
\label{fig4}
}
\end{center}
\end{figure}

   We also analyse the damped modes for $s=0$, $\ell=2$, $m=1$, which were discussed in Appendix D of Ref.~\cite{branching2} due to their peculiar behaviour. Our results, shown in Fig.~\ref{fig2} together with the zero-damped modes, are in agreement with Fig.~14 of Ref.~\cite{branching2}. In particular, we were able to resolve the spiralling behaviour of the third damped mode as $a/M$ is increased.
 
   The condition for the existence of such damped modes is given by $0 < m/(\ell+1/2) \lesssim 0.744$ and, even though this criteria was found using the eikonal limit $\ell \gg |s|$, it has been shown to be accurate even for low $\ell$. Recently, however, Hod~\cite{contro1} argued that these modes should be present even when this condition is not satisfied. Zimmerman \emph{et al.}~\cite{contro2}, using numerical techniques, were not able to find any of these modes when the condition does not hold.

   In our numerical simulations, we chose to study not only the scalar and gravitational perturbations considered in Refs.~\cite{branching1,branching2}, but also Dirac and electromagnetic fields. By varying $\ell$ and $m$, we search for damped modes of extremal Kerr black holes and perform, for the first time, a detailed analysis of their behaviour as the parameters change. Our results are presented in tables \ref{table2}-\ref{table5}: for those pairs $(\ell,m)$ for which damped modes were found, the fundamental frequency is calculated. These frequencies, and the associated separation constants, are plotted in Fig.~\ref{fig3}. In general, for fixed $m$, when $\ell$ increases, the fundamental damped modes oscillate more rapidly and become less stable. The real and imaginary parts of the associated separation constants $\lambda$ increase [following closely the spin-weighted spherical harmonics eigenvalues $(\ell-s)(\ell+s+1)$]. On the other hand, for fixed $\ell$, when $m$ increases, the fundamental damped modes also oscillate more rapidly, but become more stable. The real and imaginary parts of the separation constants $\lambda$ both decrease.
   
    Our findings are summarized in Fig.~\ref{fig4}, where we indicate which triplets $(s,\ell,m)$ allow the existence of damped modes. In particular, for $s=0$ and $s=-2$, our results are in complete agreement with Fig.~I of Ref.~\cite{branching1}. Therefore, regarding the controversy surrounding the unrestricted existence of damped modes, our results are in agreement with Refs.~\cite{branching1, branching1, contro2}. In particular, following the discussion in Ref.~\cite{contro2}, we have searched around $M\omega = m/2 + (0.162-i0.035)e^{-1.532n}$ for possible solutions of eqs.~\eqref{cond1} and \eqref{condang} which violate $0 < m/(\ell+1/2) \lesssim 0.744$ when $s=-2$, $\ell=m=2$, but no damped QNM solution was found.

%%% Damped mode tables - start

\begingroup
\squeezetable
\begin{table*}
\centering \caption{Damped modes for scalar perturbations of an extremal Kerr black hole. For a given set of parameters $\ell$ and $m$, the fundamental frequency is given. Asterisks indicate pairs $(\ell,m)$ for which no damped modes were found.} \vskip 2pt
\begin{tabular}{c|c|c|c|c|c|c|c|c|c|c|c}
\hline 
$m \backslash \ell$ & $0$ & $1$  & $2$ &$3$ & $4$ & $5$ & $6$ & $7$ & $8$ & $9$ & $10$  \\ \hline
 \multirow{2}{*}{$0$}  & $0.110245$  & $0.314986$ & $0.524122$ & $0.733303$ & $0.942564$ & $1.151863$ & $1.361184$ & $1.570518$ & $1.779860$ & $1.989208$ & $2.198560$   \\
  & $- 0.089433$i   & $ - 0.081714$i & $- 0.081323$i   & $-0.081168$i & $ - 0.081104$i   & $ - 0.081071$i & $ - 0.081052$i   & $ - 0.081040$i & $ - 0.081032$i & $ - 0.081026$i & $ - 0.081022$i  \\          
\hline 
\multirow{2}{*}{$1$}  & \X & $*$ & $0.664311 $ & $0.861758$ & $1.065265$ & $1.271142$ & $1.478194$ & $1.685913$ & $1.894048$ & $2.102459$ & $2.311064$    \\
  &    & $*$ &  $- 0.056054$i & $- 0.066005$i   & $ - 0.070348$i & $ - 0.072761$i   & $- 0.074289$i & $- 0.075343$i   & $ - 0.076112$i & $- 0.076698$i & $ - 0.077159$i   \\          
\hline 
\multirow{2}{*}{$2$}  & \X & \X & $*$ & $1.071595 $ & $1.239028$ & $1.427553$ & $1.624331$ & $1.825248$ & $2.028545$ & $2.233339$ & $2.439136$   \\
  &    &  & $*$   & $- 0.032238$i & $- 0.050608 $i   & $- 0.059198$i & $-  0.064090$i   & $- 0.067224$i & $ - 0.069395$i & $- 0.070984$i & $- 0.072195$  \\          
\hline 
\multirow{2}{*}{$3$}  &  \X & \X & \X & $*$ & $1.520325 $ & $1.647474$ & $1.814968$ & $1.998597$ & $2.190465$ & $2.387137$ & $2.586865$   \\
  &   &  &    & $*$ & $- 0.013496$i   & $-0.036397$i & $ - 0.048386$i   & $ - 0.055488$i & $ -0.060130$i & $- 0.063381$i & $- 0.065777$i  \\          
\hline 
\multirow{2}{*}{$4$}  & \X & \X & \X & \X & $*$ & $2.000665$ & $2.081394$ & $2.223730$ & $2.391239$ & $2.571813$ & $2.760105$   \\
  &    &  &    &  & $*$   & $- 0.001598$i & $-0.023928$i   & $-0.038253$i & $- 0.047182$i & $-0.053163$i & $- 0.057414$i  \\          
\hline 
\multirow{2}{*}{$5$}  & \X & \X & \X & \X & \X & $*$ & $*$ & $2.536729$ & $2.650983 $ & $2.800129$ & $2.967644$   \\
  &   &  &  &  & & $*$ & $*$   & $-0.013517$i & $- 0.028989$i & $-0.039307$i & $ -0.046426$i  \\          
\hline 
\multirow{2}{*}{$6$}  & \X & \X & \X & \X & \X & \X & $*$ & $*$ & $3.010459$ & $3.094502$ & $3.223556$   \\
  & &   &  &  &  & & $*$   & $*$ & $- 0.005476$i & $ - 0.020713$i & $-0.031949$i  \\          
\hline 
\multirow{2}{*}{$7$}  & \X & \X & \X & \X & \X & \X & \X & $*$ & $*$ & $3.500258$ & $3.552506$   \\
  & & &   &  &  &  & & $*$ & $*$ & $- 0.000470$i & $-0.013523$i  \\          
\hline 
\multirow{2}{*}{$8$}  & \X  & \X &\X & \X  & \X &\X  & \X &\X &  $*$ & $*$ & $*$   \\
    &  & & &   &  &  &  & &  $*$ & $*$ & $*$  \\          
\hline 
\multirow{2}{*}{$9$}  & \X&\X  &\X &\X &\X   & \X &\X  & \X &\X &  $*$ & $*$   \\
  & &  & & &   &  &  &  & &  $*$ & $*$ \\          
\hline 
\multirow{2}{*}{$10$}  & \X & \X& \X &\X &\X &\X   & \X & \X & \X & \X & $*$   \\
  &  & &  & & &   &  &  &  & & $*$  \\          
\hline  
\end{tabular} \label{table2}
\end{table*}
\endgroup

\begingroup
\squeezetable
\begin{table*}
\centering \caption{Damped modes for Dirac perturbations of an extremal Kerr black hole. For a given set of parameters $\ell$ and $m$, the fundamental frequency is given. Asterisks indicate pairs $(\ell,m)$ for which no damped modes were found.} \vskip 2pt
\begin{tabular}{c|c|c|c|c|c|c|c|c|c|c|c}
\hline 
$m \backslash \ell$ & $1/2$ & $3/2$  & $5/2$ &$7/2$ & $9/2$ & $11/2$ & $13/2$ & $15/2$ & $17/2$ & $19/2$ & $21/2$  \\ \hline
 \multirow{2}{*}{$1/2$}  & $*$  & $0.473956$ & $0.682532 $ & $0.891515 $ & $1.100653$ & $1.309870$ & $1.519131$ & $1.728421$ & $1.937729$ & $2.147050$ & $2.356380$   \\
  & $*$   & $- 0.067999$i & $- 0.073384$i   & $- 0.075643$i & $-0.076873$i   & $ -0.077646$i & $- 0.078175$i   & $-0.078561$i & $-0.078854$i & $-0.079085$i & $-0.079270$i  \\          
\hline 
\multirow{2}{*}{$3/2$}  &  \X & $*$ & $0.857821 $ & $1.042609$ & $1.240161 $ & $1.442531 $ & $1.647263$ & $1.853332$ & $2.060233$ & $2.267687$ & $2.475525$   \\
  &    & $*$ & $- 0.042722$i   & $ - 0.057842$i & $- 0.064563$i   & $- 0.068312$i & $- 0.070688$i   & $-0.072325$i & $-0.073517$i & $-0.074424$i & $-0.075137$i  \\          
\hline 
\multirow{2}{*}{$5/2$}  & \X & \X & $*$ & $1.289318$ & $1.437287 $ & $1.616147 $ & $1.807036$ & $2.003966$ & $2.204377$ & $2.406983$ & $2.611063$   \\
  &    &  & $*$   & $- 0.021409$i & $- 0.042982$i   & $- 0.053541$i & $ -0.059647$i   & $-0.063589$i & $ - 0.066330$i & $-0.068339$i & $-0.069873$i  \\          
\hline 
\multirow{2}{*}{$7/2$}  &  \X & \X & \X & $*$ & $1.756205$ & $1.859807$ & $2.015114 $ & $2.191116$ & $2.377718$ & $2.570523$ & $2.767286$   \\
  &   &  &    & $*$ & $- 0.006065$i   & $-0.029637$i & $- 0.043054$i   & $-0.051180$i & $-0.056547$i & $-0.060331$i & $-0.063129$i  \\          
\hline 
\multirow{2}{*}{$9/2$}  & \X & \X & \X & \X & $*$ & $*$ & $2.305459$ & $2.433825$ & $2.592399$ & $2.766705$ & $2.950337$   \\
  &    &  &    &  & $*$   & $*$ & $ -  0.018209$i   & $-0.033353$i & $- 0.043083$i & $- 0.049690$i & $ -0.054421$i  \\          
\hline 
\multirow{2}{*}{$11/2$}  &  \X & \X & \X &\X  & \X& $*$ & $*$ & $2.770804$ & $2.869788$ & $3.008991$ & $3.169438$   \\
  &   &  &  &  & & $*$ & $*$   & $- 0.008987$i & $-  0.024586$i & $- 0.035465$i & $- 0.043098$i  \\          
\hline 
\multirow{2}{*}{$13/2$}  &\X & \X  & \X & \X & \X &\X & $*$ & $*$ & $3.253231$ & $3.321032 $ & $3.439352$   \\
  & &   &  &  &  & & $*$   & $*$ & $- 0.002392$i & $- 0.016858$i & $-0.028398$i  \\          
\hline 
\multirow{2}{*}{$15/2$}  &\X &\X & \X  & \X & \X & \X &\X & $*$ & $*$ & $*$ & $3.785962$   \\
  & & &   &  &  &  & & $*$ & $*$ & $*$ & $-0.010269$i  \\          
\hline 
\multirow{2}{*}{$17/2$}  & \X & \X& \X&  \X & \X & \X & \X & \X&  $*$ & $*$ & $*$   \\
    &  & & &   &  &  &  & &  $*$ & $*$ & $*$  \\          
\hline 
\multirow{2}{*}{$19/2$}  &\X & \X & \X& \X&  \X & \X & \X & \X &\X &  $*$ & $*$   \\
  & &  & & &   &  &  &  & &  $*$ & $*$ \\          
\hline 
\multirow{2}{*}{$21/2$}  & \X  & \X&\X  & \X&\X & \X  & \X & \X & \X & \X & $*$   \\
  &  & &  & & &   &  &  &  & & $*$  \\          
\hline  
\end{tabular}
\label{table3}
\end{table*}
\endgroup

\begingroup
\squeezetable
\begin{table*}
\centering \caption{Damped modes for electromagnetic perturbations of an extremal Kerr black hole. For a given set of parameters $\ell$ and $m$, the fundamental frequency is given. Asterisks indicate pairs $(\ell,m)$ for which no damped modes were found.} \vskip 2pt
\begin{tabular}{c|c|c|c|c|c|c|c|c|c|c|c|c}
\hline 
$m \backslash \ell$ & $1$ & $2$  & $3$ &$4$ & $5$ & $6$ & $7$ & $8$ & $9$ & $10$ & $11$  \\ \hline
 \multirow{2}{*}{$0$}  & $0.274828$  & $0.501013$ & $0.716936$ & $0.929877$ & $1.141500$ & $1.352423$ & $1.562930$ & $1.773167$ & $1.983221$ & $2.193145$ & $2.402972$   \\
  & $-0.075232$i   & $-0.079365$i & $-0.080197$i   & $-0.080523$i & $-0.080684$i   & $-0.080776$i & $-0.080833$i   & $-0.080871$i & $-0.080897$i & $-0.080916$i & $-0.080931$i  \\          
\hline 
 \multirow{2}{*}{$1$}  & $*$  & $0.642174$ & $0.845425$ & $1.052494$ & $1.260686$ & $1.469351$ & $1.678255$ & $1.887296$ & $2.096423$ & $2.305606$ & $2.514828$   \\
  & $*$   & $ - 0.051754$i & $ - 0.064382$i   & $ - 0.069497$i & $- 0.072236$i   & $- 0.073934$i & $-0.075086$i   & $-0.075918$i & $- 0.076546$i & $-0.077037$i & $-0.077431$i  \\          
\hline 
\multirow{2}{*}{$2$}  & \X  & $*$ & $1.059445 $ & $1.227322 $ & $1.417480 $ & $ 1.615647 $ & $1.817659$ & $2.021821$ & $2.227310$ & $2.433676$ & $2.640651$   \\
  &    & $*$ & $- 0.028871 $i   & $- 0.049191$i & $- 0.058427$i   & $- 0.063606$i & $-0.066892$i   & $ - 0.069153$i & $- 0.070800$i & $- 0.072050$i & $-0.073030$i  \\          
\hline 
\multirow{2}{*}{$3$}  &\X  & \X & $*$ & $1.514822 $ & $1.639194 $ & $1.807018$ & $1.991365 $ & $2.183932$ & $2.381218 $ & $2.581470$ & $2.783713$   \\
  &    &  & $*$   & $- 0.011071$i & $- 0.035160 $i   & $ - 0.047683$i & $- 0.055038$i   & $-0.059817$i & $- 0.063152$i & $ - 0.065601$i & $- 0.067472$i  \\          
\hline 
\multirow{2}{*}{$4$}  &  \X &\X & \X & $*$ & $*$ & $2.075779$ & $2.217507$ & $2.385224$ & $2.566187$ & $2.754888$ & $2.948523$   \\
  &   &  &    & $*$ & $*$   & $ - 0.022872$i & $-  0.037614$i   & $- 0.046762$i & $- 0.052868$i & $ - 0.057196$i & $- 0.060408$i  \\          
\hline 
\multirow{2}{*}{$5$}  &\X  & \X &\X  & \X& $*$ & $*$ & $2.533259$ & $2.646200$ & $2.795154$ & $2.962808$ & $3.141538$   \\
  &    &  &    &  & $*$   & $*$ & $-0.012659$i   & $ - 0.028414$i & $ - 0.038917$i & $ -0.046148$i & $-  0.051364$i  \\          
\hline 
\multirow{2}{*}{$6$}  & \X  & \X & \X & \X &\X & $*$ & $*$ & $3.008760 $ & $3.090940$ & $3.219481$ & $3.372458$   \\
  &   &  &  &  & & $*$ & $*$   & $- 0.004854$i & $ - 0.020205$i & $ - 0.031589$i & $-0.039723$i  \\          
\hline 
\multirow{2}{*}{$7$}  & \X & \X  & \X & \X & \X & \X & $*$ & $*$ & $*$ & $3.549997$ & $3.656826$   \\
  & &   &  &  &  & & $*$   & $*$ & $*$ & $-0.013087$i & $-0.024839$i  \\          
\hline 
\multirow{2}{*}{$8$}  & \X & \X & \X  & \X & \X & \X & \X & $*$ & $*$ & $*$ & $4.021952$   \\
  & & &   &  &  &  & & $*$ & $*$ & $*$ & $- 0.007176$i  \\          
\hline 
\multirow{2}{*}{$9$}  &\X  & \X&\X & \X  & \X &\X  & \X &\X &  $*$ & $*$ & $*$   \\
    &  & & &   &  &  &  & &  $*$ & $*$ & $*$  \\          
\hline 
\multirow{2}{*}{$10$}  &\X & \X &\X & \X& \X  & \X & \X & \X &\X &  $*$ & $*$   \\
  & &  & & &   &  &  &  & &  $*$ & $*$ \\          
\hline 
\multirow{2}{*}{$11$}  & \X & \X&  \X& \X& \X& \X  & \X & \X & \X & \X & $*$   \\
  &  & &  & & &   &  &  &  & & $*$  \\          
\hline  
\end{tabular}
\label{table4}
\end{table*}  
\endgroup

\begingroup
\squeezetable
\begin{table*}
\centering \caption{Damped modes for gravitational perturbations of an extremal Kerr black hole. For a given set of parameters $\ell$ and $m$, the fundamental frequency is given. Asterisks indicate pairs $(\ell,m)$ for which no damped modes were found.} \vskip 2pt
\begin{tabular}{c|c|c|c|c|c|c|c|c|c|c|c|c}
\hline 
$m \backslash \ell$ & $2$ &$3$ & $4$ & $5$ & $6$ & $7$ & $8$ & $9$ & $10$ & $11$ & $12$  \\ \hline
 \multirow{2}{*}{$0$}  & $0.425145$  & $0.665132$ & $0.890510$ & $1.109688$ & $1.325702$ & $1.539879$ & $1.752892$ & $1.965120$ & $2.176794$ & $2.388061$ & $2.599020$   \\
  & $-0.071806$i   & $-0.076735$i & $-0.078575$i   & $-0.079431$i & $-0.079900 $i   & $-0.080185$i & $- 0.080372$i   & $-0.080501$i & $-0.080594$i & $-0.080663$i & $-0.080716$i  \\          
\hline 
 \multirow{2}{*}{$1$}  & $0.581433$  & $0.795283 $ & $1.013254$ & $1.228731$ & $1.442442$ & $1.655025$ & $1.866861$ & $2.078182$ & $2.289134$ & $2.499811$ & $2.710280$   \\
  & $-0.038255$i   & $- 0.058965$i & $- 0.066719$i   & $ - 0.070562$i & $ - 0.072817$i   & $-0.074287$i & $-0.075318$i   & $-0.076079$i & $-0.076663$i & $-0.077124$i & $-0.077498$i  \\          
\hline 
 \multirow{2}{*}{$2$}  & $*$  & $1.028553$ & $1.192475$ & $1.387017$ & $1.589346$ & $1.794694$ & $2.001500$ & $2.209111$ & $2.417208$ & $2.625619$ & $2.834244$   \\
  & $*$   & $- 0.018572$i & $- 0.044695$i   & $- 0.056001$i & $-0.062097$i   & $- 0.065865$i & $ - 0.068408$i   & $-0.070235$i & $-0.071607$i & $-0.072673$i & $-0.073526$i  \\          
\hline 
\multirow{2}{*}{$3$}  & \X  & $*$ & $1.503222$ & $1.615078$ & $1.783228$ & $1.969595$ & $2.164245 $ & $2.363378$ & $2.565215$ & $2.768810$ & $2.973612$   \\
  &    & $*$ & $- 0.004371$i   & $- 0.031331$i & $- 0.045508$i   & $ - 0.053651$i & $-  0.058858$i   & $ - 0.062450$i & $-0.065065$i & $-0.067049$i & $-0.068604$i  \\          
\hline 
\multirow{2}{*}{$4$}  & \X & \X & $*$ & $*$ & $2.059795$ & $2.199039$ & $2.367203$ & $2.549284$ & $2.739199$ & $2.933980$ & $3.131996$   \\
  &    &  & $*$   & $*$ & $ -0.019667$i   & $-0.035658$i & $ - 0.045480$i   & $- 0.051968$i & $- 0.056531$i & $-0.059897$i & $ - 0.062473$i  \\          
\hline 
\multirow{2}{*}{$5$}  &  \X & \X&  \X& $*$ & $*$ & $2.523730$ & $2.632115$ & $2.780307$ & $2.948312$ & $3.127738$ & $3.314388$   \\
  &   &  &    & $*$ & $*$   & $-0.010112$i & $-0.026668$i   & $-0.037732$i & $- 0.045302$i & $ - 0.050732$i & $-0.054791$i  \\          
\hline 
\multirow{2}{*}{$6$}  & \X & \X & \X &\X & $*$ & $*$ & $3.004516$ & $3.080548 $ & $3.207368 $ & $3.360070$ & $3.527952 $   \\
  &    &  &    &  & $*$   & $*$ & $- 0.003097$i   & $- 0.018673$i & $- 0.030500$i & $- 0.038929$i & $- 0.045110$i  \\          
\hline 
\multirow{2}{*}{$7$}  & \X  & \X & \X & \X & \X& $*$ & $*$ & $*$ & $3.542772$ & $3.647099$ & $3.783490$   \\
  &   &  &  &  & & $*$ & $*$   & $*$ & $- 0.011784$i & $- 0.023846$i & $- 0.032898$i  \\          
\hline 
\multirow{2}{*}{$8$}  &\X & \X  & \X & \X & \X &\X & $*$ & $*$ & $*$ & $4.017480$ & $4.098411$   \\
  & &   &  &  &  & & $*$   & $*$ & $*$ & $0.006130$ & $-0.017821$i  \\          
\hline 
\multirow{2}{*}{$9$}  &\X &\X & \X  & \X & \X &\X  &\X & $*$ & $*$ & $*$ & $4.503570$   \\
  & & &   &  &  &  & & $*$ & $*$ & $*$ & $-0.001937$i  \\          
\hline 
\multirow{2}{*}{$10$}  & \X & \X&\X & \X  & \X & \X & \X &\X &  $*$ & $*$ & $*$   \\
    &  & & &   &  &  &  & &  $*$ & $*$ & $*$  \\          
\hline 
\multirow{2}{*}{$11$}  &\X & \X &\X & \X &\X   &\X  &\X  & \X  &\X &  $*$ & $*$   \\
  & &  & & &   &  &  &  & &  $*$ & $*$ \\          
\hline 
\multirow{2}{*}{$12$}  & \X & \X& \X & \X& \X& \X  & \X &\X  & \X & \X & $*$   \\
  &  & &  & & &   &  &  &  & & $*$  \\          
\hline  
\end{tabular}
\label{table5}
\end{table*}  
\endgroup
%  

%%% Damped mode tables - end

 Finally, for $m<0$ (counter-rotating modes), only damped modes are allowed. Once again, we compare our numerical results with the QNMs of a near extremal Kerr black hole for several parameters $s$, $\ell$, and $m$, as shown in table \ref{table6}. Similarly to the $m=0$ case, we find excellent agreement between the results. 
 
\begin{table*}
\centering \caption{Comparison of the QNM frequencies $M\omega$ and the associated separation constants $\lambda$ between a near extremal ($a/M=0.999$) and an extremal Kerr black hole for $m<0$. For each set of parameters $s$, $\ell$, and $m<0$, we have calculated the least damped mode ($n=0$) and the the first overtone ($n=1$).} \vskip 2pt
\begin{tabular}{@{}c|c|c|c|cccc|cccc@{}}
\hline 
$s$&$\ell$&$m$&$n$&\multicolumn{4}{c|}{$a/M=0.999$}&\multicolumn{4}{c}{$a/M=1$}\\ \hline
   &      &   & & $\text{Re}(M\omega)$     &$\text{Im}(M\omega)$ & $\text{Re}(\lambda)$ &$\text{Im}(\lambda)$ & $\text{Re}(M\omega)$ &$\text{Im}(M\omega)$ & $\text{Re}(\lambda)$ &$\text{Im}(\lambda)$  \\
 $0$    & $1$  & $-1$ & $0$ & $0.239462$ & $-0.093828$  &  $1.990310$ & $0.008989$  & $0.239424$ & $-0.093821$ & $1.990294$ & $0.009005$  \\
  $0$    & $1$  & $-1$ & $1$  & $0.201620$ & $-0.300295$  &  $2.009941$ & $0.024115$  & $0.201573$ & $-0.300280$ & $2.009963$ & $0.024156$ \\
 $-\frac{1}{2}$    & $\frac{1}{2}$  & $-\frac{1}{2}$ & $0$   & $0.159112$ & $-0.093385$  &  $1.046242$ & $-0.019324$  & $0.159089$ & $-0.093376$ & $1.046236$ & $-0.019324$  \\
 $-\frac{1}{2}$    & $\frac{1}{2}$  & $-\frac{1}{2}$ & $1$ & $0.116224$ & $-0.311875$  &  $1.071028$ & $-0.073697$  & $0.116178$ & $-0.311858$ & $1.071013$ & $-0.073704$ \\           
 $-1$    & $1$  & $-1$ & $0$   & $0.204380$ & $-0.091352$  &  $2.186087$ & $-0.071537$  & $0.204349$ & $-0.091348$ & $2.186231$ & $-0.071589$  \\
$-1$    & $1$  & $-1$ & $1$ & $0.158716$ & $-0.299037$  &  $2.190971$ & $-0.246458$  & $0.158677$ & $-0.299029$ & $2.191157$ & $-0.246657$  \\           
 $-2$    & $2$  & $-1$ & $0$  & $0.343864$ & $-0.083401$  &  $4.395467$ & $-0.079325$  & $0.343862$ & $-0.083384$ & $4.395797$ & $-0.079358$  \\
 $-2$    & $2$  & $-1$ & $1$ & $0.316309$ & $-0.257080$  &  $4.400545$ & $-0.251561$  & $0.316305$ & $-0.257024$ & $4.400904$ & $-0.251671$ \\         
 $-2$    & $2$  & $-2$ & $0$   & $0.291609$ & $-0.088028$  &  $4.721185$ & $-0.198122$  & $0.291553$ & $-0.088026$ & $4.721727$ & $-0.198286$  \\
 $-2$    & $2$  & $-2$ & $1$  & $0.250213$ & $-0.276742$  &  $4.674323$ & $-0.637396$  & $0.250146$ & $-0.276738$ & $4.674850$ & $-0.637954$ \\           
\hline 
\end{tabular}
\label{table6}
\end{table*}

%%%%%%%%%%%%%%%%%%%%%%%
%%%%%%%%%%%%%%%%%%%%%%%
%%%%%%%%%%%%%%%%%%%%%%%%

\subsection{Extremal RN black holes} 

\begin{figure*}
\begin{center}
\begin{tabular}{cc}
\includegraphics[width=8.6cm]{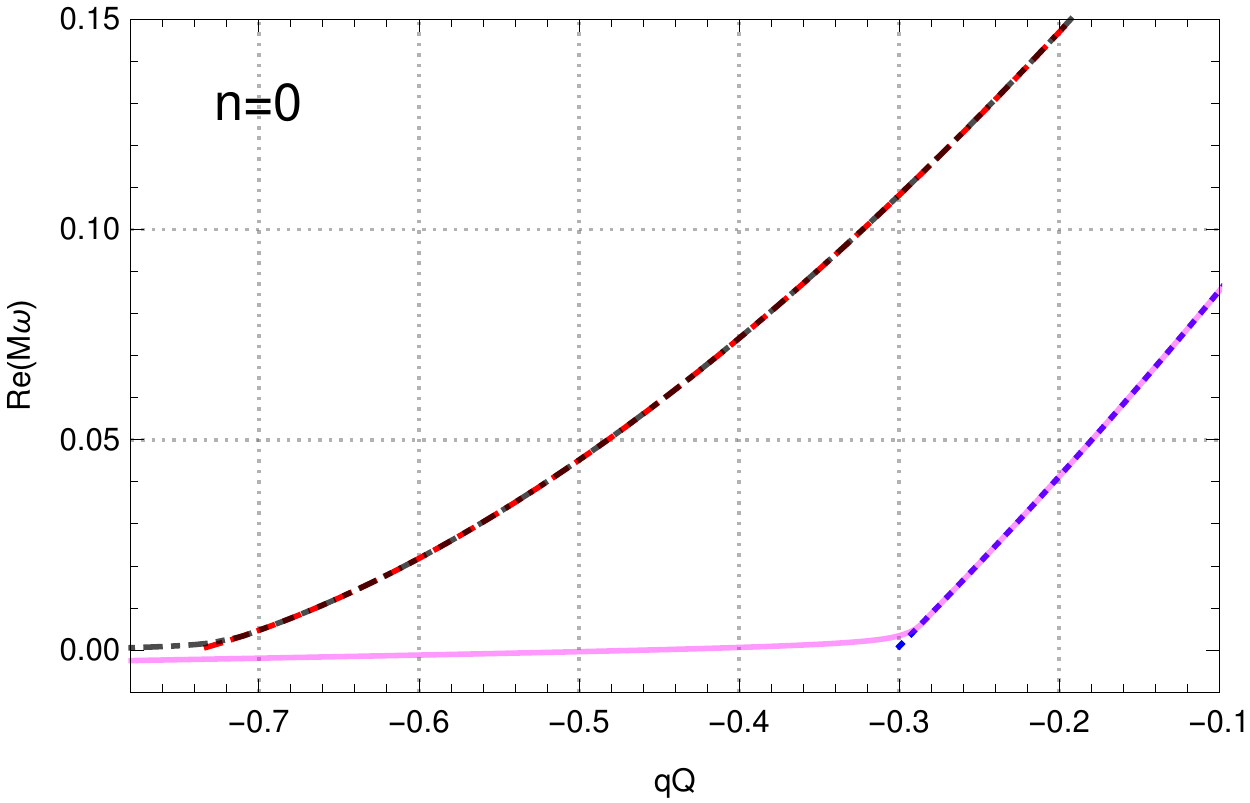}&
\includegraphics[width=8.6cm]{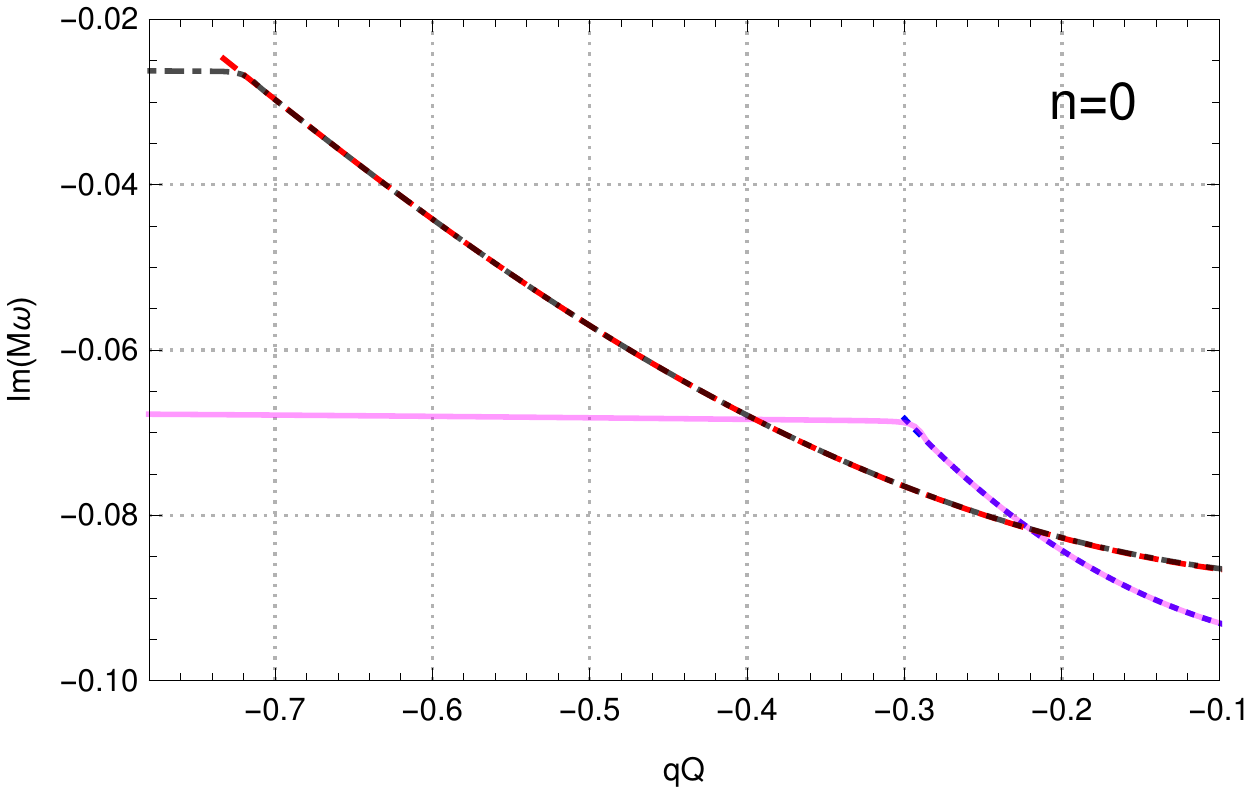}\\
\includegraphics[width=8.6cm]{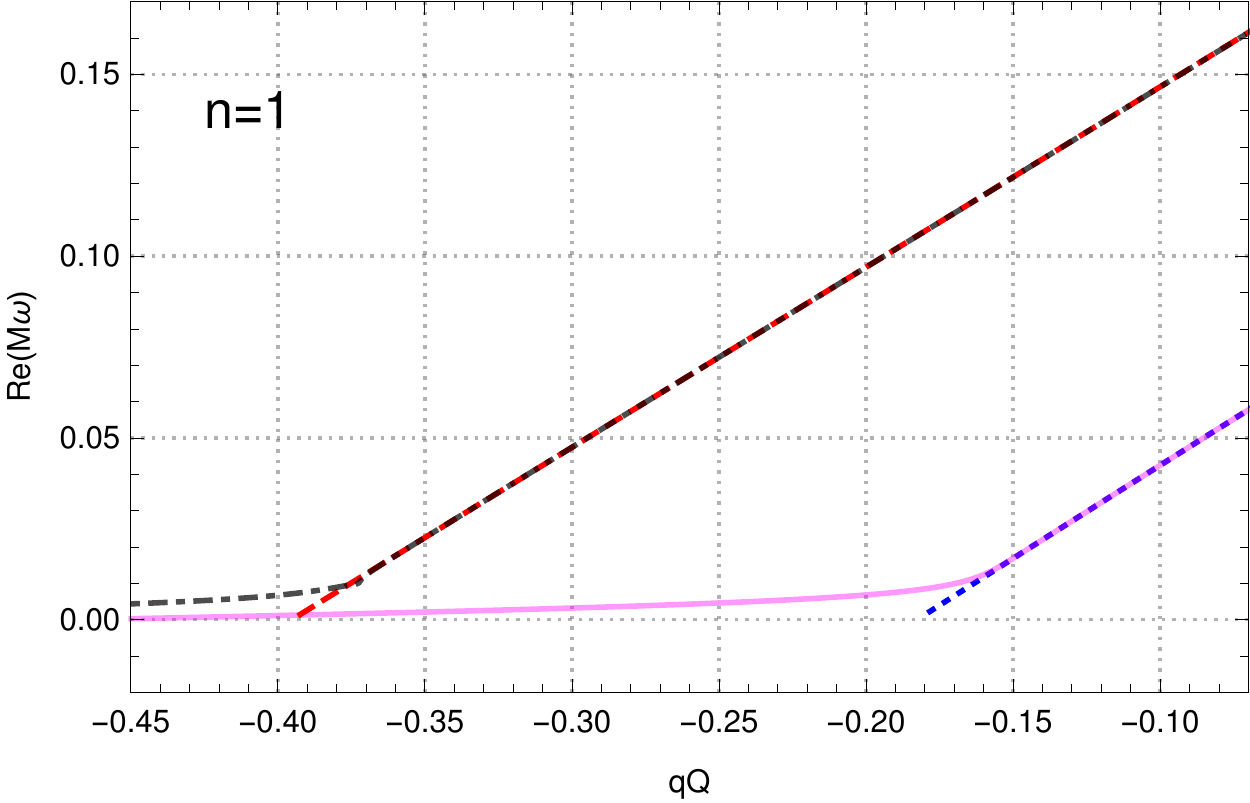}&
\includegraphics[width=8.6cm]{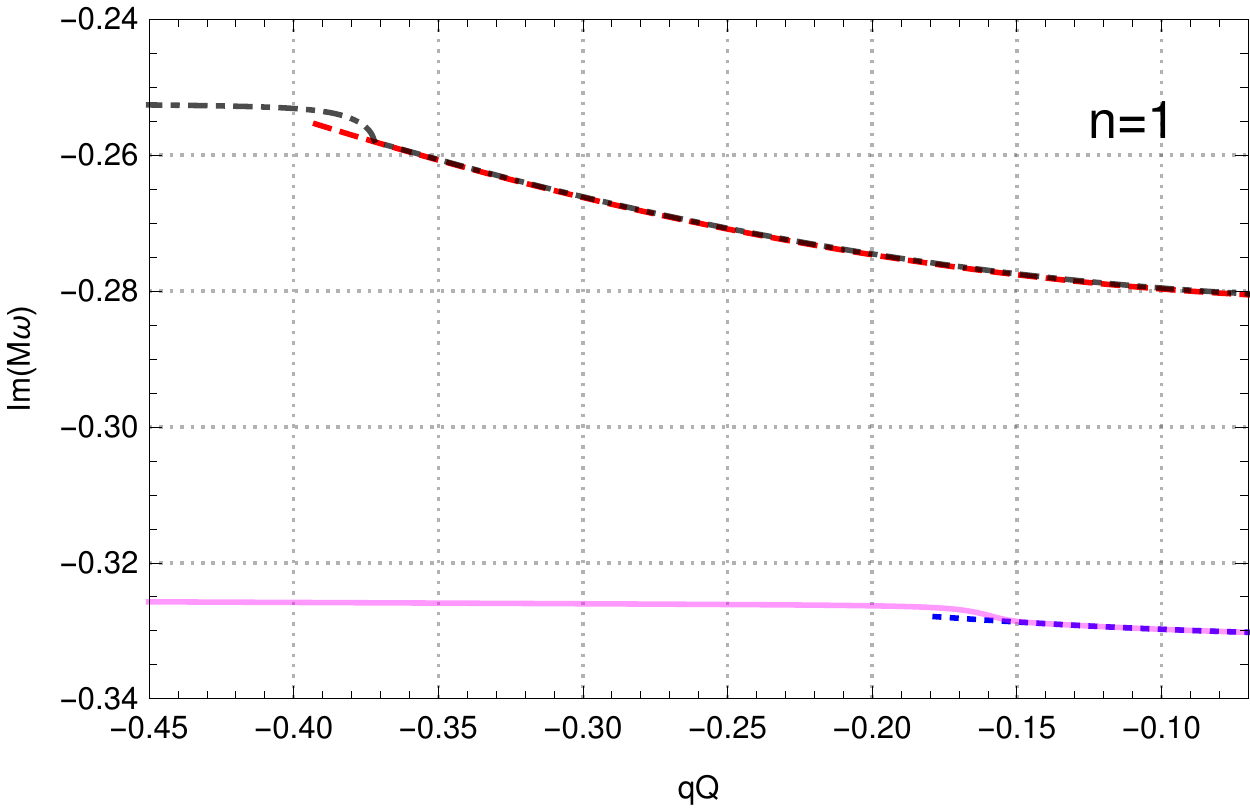}
\end{tabular}
\caption{(Colors online.) Real and imaginary parts of the QNMs for negative values of $qQ$. The blue (dotted) and the red (dashed) curves correspond, respectively, to scalar ($s=j=0$) and Dirac ($s=-1/2$, $\ell=1/2$) fields around a near extremal ($Q/M=0.999$) RN black hole, while the magenta (solid) and the black (dot-dashed) curves correspond to scalar ($s=\ell=0$) and Dirac ($s=-1/2$, $\ell=1/2$) fields around an extremal RN black hole. The top and bottom plots correspond, respectively, to fundamental QNMs and its first overtones. Note the critical value of $qQ$ below which the quasinormal branches disappear for near extremal black holes.
\label{fig5}
}
\end{center}
\end{figure*}

\begin{figure*}
\begin{center}
\begin{tabular}{cc}
\includegraphics[width=8.6cm]{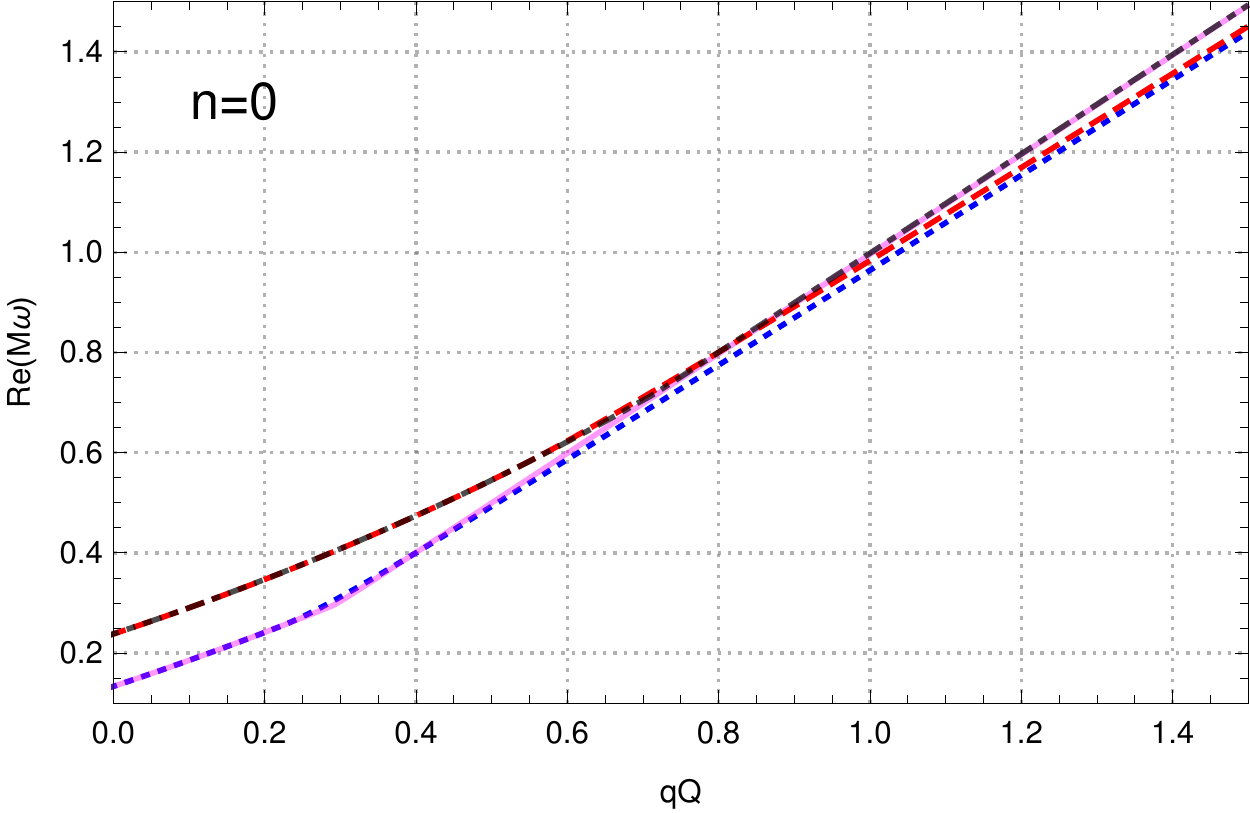}&
\includegraphics[width=8.6cm]{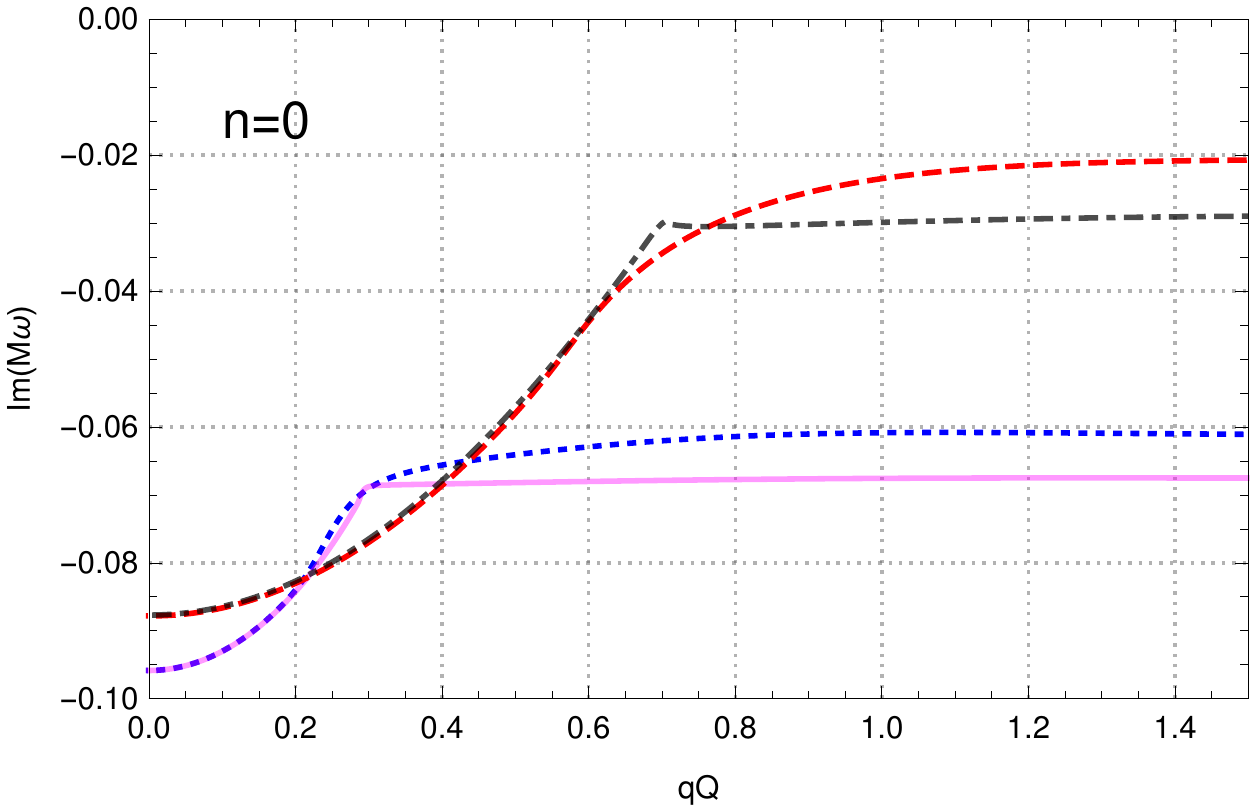}\\
\includegraphics[width=8.6cm]{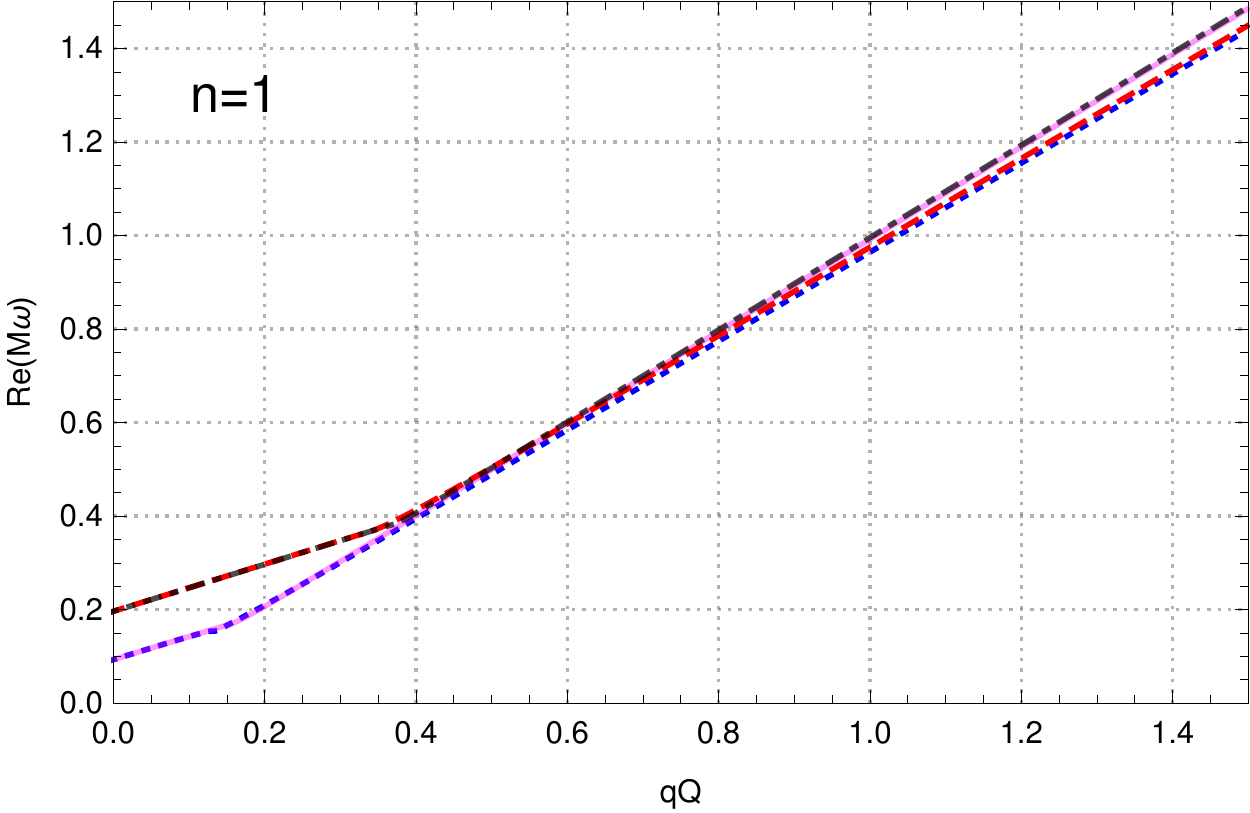}&
\includegraphics[width=8.6cm]{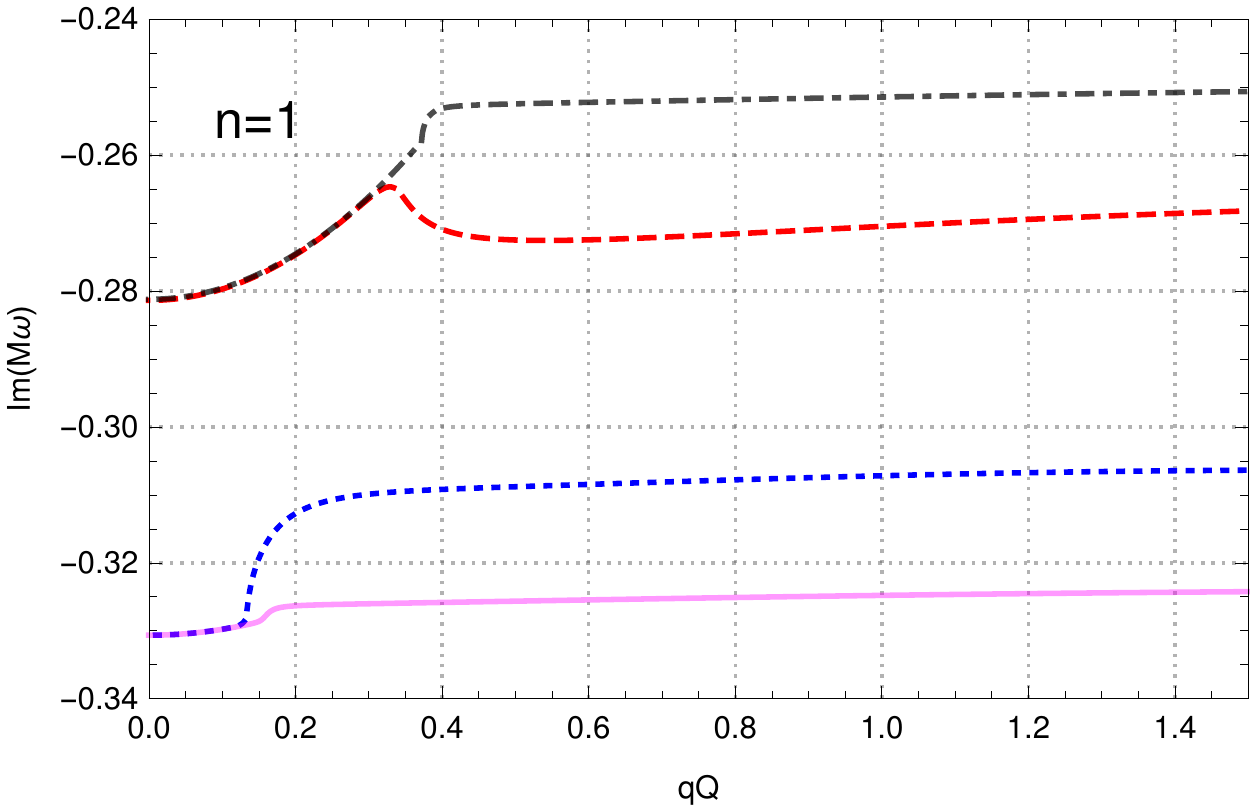}
\end{tabular}
\caption{(Colors online.) Real and imaginary parts of the quasinormal frequencies for positive value of $qQ$. The blue (dotted) and the red (dashed) curves correspond, respectively, to scalar ($s=j=0$) and Dirac ($s=-1/2$, $\ell=1/2$) fields around a near extremal ($Q/M=0.999$) RN black hole, while the magenta (solid) and the black (dot-dashed) curves correspond to scalar ($s=\ell=0$) and Dirac ($s=-1/2$, $\ell=1/2$) fields around an extremal RN black hole. The top and bottom plots correspond, respectively, to fundamental QNMs and its first overtones. Note the difference between extremal and near extremal black holes as the electromagnetic interaction $qQ$ is increased.
\label{fig6}
}
\end{center}
\end{figure*}

The first (and, until now, only) work to discuss the implementation of a continued fraction technique to extremal black holes was Ref.~\cite{onozawa}. As explained before, they considered only neutral perturbations of extremal RN black holes and their implementation relied on the fact that the coefficient $\delta_n$ in eq.~\eqref{5term} vanishes for scalar, electromagnetic, and gravitational fields (spin-$1/2$ perturbations were not considered). Our analysis for extremal RN black holes, on the other hand, considers spin-$0$ and spin-$1/2$ charged perturbations. Furthermore, our technique works when $\delta_n \neq 0$, which is always the case for charged perturbations around an extremal black hole (and also for neutral perturbations around an extremal Kerr black hole). 

Contrary to Kerr QNMs, RN QNMs do not depend on the azimuthal number $m$. They depend, however, on an analogous (dimensionless) quantity, $qQ$, which accounts for the electromagnetic interaction between the black hole and the perturbation field. Additionally, while $m$ is a discrete parameter, $qQ$ is continuous. Having this in mind, we now analyse the behaviour of the quasinormal frequencies for an extremal RN black hole as the electromagnetic interaction term is varied. 

   In our simulations, we start from $|qQ| \ll 1$ and track the quasinormal modes as $qQ$ increases. Our results, presented in tables \ref{table7} and \ref{table8} for several combinations of the parameters $s$, $\ell$, and $n$, demonstrate that the quasinormal frequencies for extremal RN black holes are in excellent agreement with the QN frequencies of a near extremal black hole for $qQ \lesssim$ $0.1$. In particular, for $s=0$, as $qQ \rightarrow 0$, our results approach the values calculated in Ref.~\cite{onozawa} for neutral perturbations. As $|qQ|$ is increased, however, we find unexpected results.

\begingroup
\squeezetable 
\begin{table*}
\centering
\caption{The QNM frequencies $M\omega$ of an extremal RN black hole for positive values of the electromagnetic interaction parameter $qQ$.}  \vskip 2pt
\begin{tabular}{@{}c|c|c|cc|cc|cc|cc|cc|cc@{}}
\hline 
$s$&$\ell$&$n$ &\multicolumn{4}{c|}{$qQ=0.001$}  & \multicolumn{4}{c|}{$qQ=0.01$} &\multicolumn{4}{|c}{$qQ=0.1$} \\ \hline
 &   &   & \multicolumn{2}{c|}{$Q/M=0.999$} & \multicolumn{2}{c|}{$Q/M=1$} & \multicolumn{2}{c|}{$Q/M=0.999$} & \multicolumn{2}{c|}{$Q/M=1$} & \multicolumn{2}{c|}{$Q/M=0.999$} & \multicolumn{2}{|c}{$Q/M=1$} \\  \hline
 &   &   & $\text{Re}(M\omega)$     &$\text{Im}(M\omega)$ & $\text{Re}(M\omega)$     &$\text{Im}(M\omega)$ & $\text{Re}(M\omega)$ &$\text{Im}(M\omega)$ & $\text{Re}(M\omega)$     &$\text{Im}(M\omega)$ & $\text{Re}(M\omega)$     &$\text{Im}(M\omega)$ & $\text{Re}(M\omega)$     &$\text{Im}(M\omega)$  \\
$0$ &  $0$ & $0$   & $0.133959$ & $-0.095843$  &  $0.133959$ & $ -0.095844$  & $0.138479$ & $ -0.095815 $ & $ 0.138478 $ & $ - 0.095816 $ & $ 0.185412 $ & $  -0.093009 $ & $ 0.185411$ & $-0.093018 $  \\          
$0$ &  $0$ & $1$ & $0.093464 $   & $- 0.330652$ & $0.093465$  &  $ -0.330652$ & $ 0.097959 $  & $ -0.330643$ & $ 0.097960 $ & $ -0.330644$ & $ 0.142213 $ & $  -0.329774 $ & $ 0.142459 $ & $ - 0.329781$   \\       
$0$ &  $0$ & $2$ & $0.075582$   & $ -0.588328$ & $0.075581$  &  $ -0.588326$ & $0.080081 $  & $  -0.588326$ & $ 0.080079 $ & $  -0.588323$ & $ 0.126173 $ & $ -0.589479$ & $ 0.124830  $ & $  -0.588012$   \\
$0$ &  $1$ & $0$ & $0.377854$   & $  -0.089529$ & $0.378142  $  &  $  -0.089384$ & $ 0.382363 $  & $ - 0.089528$ & $ 0.382657  $ & $  -0.089379$ & $ 0.428778 $ & $- 0.089044$ & $ 0.429135 $ & $ -0.088838$   \\
$0$ &  $1$ & $1$ & $0.348615$   & $  -0.276440$ & $0.348680$  &  $ -0.276139$ & $ 0.353125 $  & $ - 0.276434$ & $ 0.353187 $ & $ -0.276128$ & $ 0.398835  $ & $ - 0.275353$ & $ 0.398861 $ & $ -0.274988$   \\
$0$ &  $1$ & $2$ & $0.299098 $   & $ - 0.486659$ & $0.298958 $  &  $  -0.486435$ & $ 0.303602 $  & $ - 0.486649$ & $ 0.303458 $ & $ -0.486425$ & $0.348584  $ & $ - 0.485725$ & $ 0.348396 $ & $  -0.485507 $   \\
$0$ &  $2$ & $0$ & $0.626504 $   & $ -0.088912$ & $0.627073 $  &  $ - 0.088748$ & $ 0.631006 $  & $ - 0.088913$ & $ 0.631582  $ & $ - 0.088746$ & $ 0.676875  $ & $ - 0.088747$ & $ 0.677534  $ & $ - 0.088542$   \\
$0$ &  $2$ & $1$ & $0.608278 $   & $ - 0.269526$ & $0.608671$  &  $ - 0.269093$ & $ 0.612783 $  & $  -0.269528$ & $ 0.613179 $ & $ -0.269087$ & $ 0.658498  $ & $  -0.269068$ & $ 0.658918 $ & $ -0.268547$   \\
$0$ &  $2$ & $2$ & $0.573255 $   & $- 0.458754$ & $0.573373 $  &  $ - 0.458202$ & $ 0.577762 $  & $  -0.458754$ & $ 0.577877  $ & $ - 0.458195$ & $0.623207  $ & $ - 0.458121$ & $ 0.623285  $ & $ - 0.457499$   \\
%FERMION%
%
$-\frac{1}{2}$ &  $\frac{1}{2}$ & $0$ & $0.238550$   & -$0.087811$ & $0.238682$  &  $ -0.087685$ & $ 0.243068 $  & $ - 0.087805$ & $ 0.243204 $ & $ -0.087672$ & $ 0.290216 $ & $ - 0.086637$ & $ 0.290379 $ & $ -0.086432$   \\          
$-\frac{1}{2}$ &  $\frac{1}{2}$ & $1$ & $0.196977$   & $-0.281368$ & $0.196896$  &  $ -0.281211$ & $ 0.201484$  & $ -0.281352$ & $ 0.201397 $ & $ -0.281194$ & $ 0.246685 $ & $ -0.279700$ & $ 0.246546 $ & $ -0.279545$   \\  
$-\frac{1}{2}$ &  $\frac{1}{2}$ & $2$ & $0.146203$   & $-0.515626$ & $0.146077$  &  $ -0.515585$ & $ 0.150701$  & $ -0.515617$ & $ 0.150574 $ & $ -0.515578$ & $ 0.195347 $ & $ -0.514889$ & $ 0.195209 $ & $ -0.514871$   \\ 
$-\frac{1}{2}$ &  $\frac{3}{2}$ & $0$ & $0.494176$   & $-0.088402$ & $0.494613$  &  $ -0.088240$ & $ 0.498681$  & $ -0.088403$ & $ 0.499125 $ & $ -0.088237$ & $ 0.544802 $ & $ -0.088128$ & $ 0.545324 $ & $ -0.087913$   \\ 
$-\frac{1}{2}$ &  $\frac{3}{2}$ & $1$ & $0.471418$   & $-0.269611$ & $0.471647$  &  $ -0.269213$ & $ 0.475926$  & $ -0.269611$ & $ 0.476154 $ & $ -0.269205$ & $ 0.521728 $ & $ -0.268898$ & $ 0.521952$ & $ -0.268408$   \\ 
$-\frac{1}{2}$ &  $\frac{3}{2}$ & $2$ & $0.429052$   & $-0.464470$ & $0.429001$  &  $ -0.464039$ & $ 0.433560$  & $ -0.464465$ & $ 0.433504 $ & $ -0.464030$ & $ 0.478881 $ & $ -0.463620$ & $ 0.478768 $ & $ -0.463151$   \\ 
$-\frac{1}{2}$ &  $\frac{5}{2}$ & $0$ & $0.745878$   & $-0.088496$ & $0.746584$  &  $ -0.088327$ & $0.750378$  & $ -0.088498$ & $ 0.751092 $ & $ -0.088325$ & $ 0.796105 $ & $ -0.088385$ & $ 0.796906 $ & $ -0.088180$   \\ 
$-\frac{1}{2}$ &  $\frac{5}{2}$ & $1$  & $0.730535$ & $0.267422$   & $0.731087$  &  $ -0.266956$ & $0.735038$  & $ -0.267425$ & $ 0.735594 $ & $ -0.266952$ & $ 0.780680 $ & $ -0.267105$ & $ 0.781278 $ & $ -0.266553$   \\ 
$-\frac{1}{2}$ &  $\frac{5}{2}$ & $2$ & $0.700602$   & $-0.452296$ & $0.700890$  &  $ -0.451648$ & $ 0.705107$  & $ -0.452298$ & $ 0.705394 $ & $ -0.451643$ & $0.750590 $ & $ -0.451822$ & $ 0.750861 $ & $ -0.451087$   \\ 
\hline 
\end{tabular}
\label{table7}
\end{table*}  
\endgroup

\begingroup
\squeezetable
\begin{table*}
\centering \caption{The QNM frequencies $M\omega$ of an extremal RN black hole for negative values of the electromagnetic interaction parameter $qQ$.} \vskip 2pt
\begin{tabular}{@{}c|c|c|cc|cc|cc|cc|cc|cc@{}}
\hline 
$s$&$\ell$&$n$ &\multicolumn{4}{c|}{$qQ=-0.001$}  & \multicolumn{4}{c|}{$qQ=-0.01$} &\multicolumn{4}{|c}{$qQ=-0.1$} \\ \hline
 &   &   & \multicolumn{2}{c|}{$Q/M=0.999$} & \multicolumn{2}{c|}{$Q/M=1$} & \multicolumn{2}{c|}{$Q/M=0.999$} & \multicolumn{2}{c|}{$Q/M=1$} & \multicolumn{2}{c|}{$Q/M=0.999$} & \multicolumn{2}{|c}{$Q/M=1$} \\  \hline
 &   &   & $\text{Re}(M\omega)$     &$\text{Im}(M\omega)$ & $\text{Re}(M\omega)$     &$\text{Im}(M\omega)$ & $\text{Re}(M\omega)$ &$\text{Im}(M\omega)$ & $\text{Re}(M\omega)$     &$\text{Im}(M\omega)$ & $\text{Re}(M\omega)$     &$\text{Im}(M\omega)$ & $\text{Re}(M\omega)$     &$\text{Im}(M\omega)$  \\
$0$ &  $0$ & $0$   & $0.132959$ & $- 0.095843$ & $ 0.132959$ & $ -0.095844 $ &  $ 0.128479$ & $ -0.095815$  & $ 0.128478$ & $ -0.095816 $ & $ 0.085413$ & $ -0.093016 $  &  $ 0.085411$ & $-0.093018 $  \\          
$0$ &  $0$ & $1$ & $0.092465$   & $-0.330652$ & $0.092465$  &  $ -0.330652$ & $ 0.087960$  & $ -0.330643$ & $ 0.087960 $ & $ -0.330644$ & $0.042459 $ & $ -0.329779$ & $ 0.042459 $ & $ -0.329781$   \\       
$0$ &  $0$ & $2$ & $0.074582$   & $-0.588327$ & $ 0.074581$  & $ -0.588326$ & $0.070079$  &  $ -0.588324$  & $ 0.070079 $ & $ -0.588323$ & $ 0.024829 $ & $ -0.588012$ & $ 0.024830 $ & $ -0.588012$   \\
$0$ &  $1$ & $0$ & $0.376855$   & $-0.089528$ & $0.377142$  &  $ -0.089384$ & $0.372376$  & $ -0.089518$ & $ 0.372657 $ & $ -0.089379$ & $ 0.328906 $ & $ -0.088937$ & $ 0.329135 $ & $ -0.088838$   \\
$0$ &  $1$ & $1$ & $0.347615$   & $-0.276439$ & $ 0.347680$  & $ -0.276139$ & $0.343119$  &  $ -0.276422$  & $ 0.343187 $ & $ -0.276128$ & $ 0.298773 $ & $ -0.275229$ & $ 0.298861 $ & $ -0.274988$   \\
$0$ &  $1$ & $2$ & $0.298097$   & $-0.486659$ & $0.297958$  &  $ -0.486435$ & $ 0.293592$  & $ -0.486649$ & $ 0.293458 $ & $ -0.486425$ & $ 0.248492 $ & $ -0.485727$ & $ 0.248396 $ & $ -0.485507$   \\
$0$ &  $2$ & $0$ & $0.625506$   & $-0.088911$ & $0.626073$  &  $ -0.088748$ & $ 0.621023$  & $ -0.088906$ & $ 0.621582 $ & $ -0.088746$ & $ 0.577046 $ & $ -0.088671$ & $ 0.577534 $ & $ -0.088542$   \\
$0$ &  $2$ & $1$ & $0.607278$   & $-0.269524$ & $0.607671$  &  $ -0.269093$ & $ 0.602788$  & $ -0.269511$ & $ 0.603179 $ & $ -0.269087$ & $0.558553 $ & $ -0.268902$ & $ 0.558918 $ & $ -0.268547$   \\
$0$ &  $2$ & $2$ & $0.572254$   & $-0.458753$ & $0.572373$  &  $ -0.458202$ & $ 0.567755$  & $ -0.458740$ & $ 0.567877 $ & $ -0.458195$ & $ 0.523138 $ & $ -0.457984$ & $ 0.523285 $ & $ -0.457499$   \\
%FERMION%
%
$-\frac{1}{2}$ &  $\frac{1}{2}$ & $0$ & $0.237550$   & $-0.087810$ & $0.237682$  &  $ -0.087685$ & $ 0.233075$  & $ -0.087792$ & $ 0.233204 $ & $ -0.087672$ & $ 0.190278 $ & $ -0.086505$ & $ 0.190379 $ & $ -0.086432$   \\          
$-\frac{1}{2}$ &  $\frac{1}{2}$ & $1$ & $0.195976$   & $-0.281368$ & $ 0.195896$  & $ -0.281211$ & $0.191473$  &  $ -0.281350$ & $ 0.191397 $ & $ -0.281194$ & $ 0.146583 $ & $ -0.279689$ & $ 0.146546 $ & $ -0.279545$   \\  
$-\frac{1}{2}$ &  $\frac{1}{2}$ & $2$ & $0.145203$   & $-0.515627$ & $0.145077$  &  $ -0.515585$ & $ 0.140698$  & $ -0.515621$ & $ 0.140574 $ & $ -0.515578$ & $ 0.095320 $ & $ -0.514930$ & $ 0.095209 $ & $ -0.514871$   \\ 
$-\frac{1}{2}$ &  $\frac{3}{2}$ & $0$ & $0.493178$   & $-0.088401$ & $0.493613$  &  $ -0.088240$ & $ 0.488697$  & $ -0.088394$ & $ 0.489125 $ & $ -0.088237$ & $ 0.444962 $ & $ -0.088034$ & $ 0.445324 $ & $ -0.087913$   \\ 
$-\frac{1}{2}$ &  $\frac{3}{2}$ & $1$ & $0.470418$   & $-0.269610$ & $0.470647$  &  $ -0.269213$ & $ 0.465926$  & $ -0.269594$ & $ 0.466154 $ & $ -0.269205$ & $ 0.421728 $ & $ -0.268728$ & $ 0.421952 $ & $ -0.268408$   \\ 
$-\frac{1}{2}$ &  $\frac{3}{2}$ & $2$ & $0.428051$   & $-0.464469$ & $0.428001$  &  $ -0.464039$ & $ 0.423549$  & $ -0.464457$ & $ 0.423504 $ & $ -0.464030$ & $ 0.378770 $ & $ -0.463539$ & $ 0.378768 $ & $ -0.463151$   \\ 
$-\frac{1}{2}$ &  $\frac{5}{2}$ & $0$ & $0.744880$   & $-0.088495$ & $0.745584$  &  $ -0.088327$ & $ 0.740396$  & $ -0.088491$ & $ 0.741092 $ & $ -0.088325$ & $ 0.696286 $ & $ -0.088319$ & $ 0.696906 $ & $ -0.088180$   \\ 
$-\frac{1}{2}$ &  $\frac{5}{2}$ & $1$ & $0.729536$   & $-0.267420$ & $0.730087$  &  $ -0.266956$ & $ 0.725047$  & $ -0.267409$ & $ 0.725594 $ & $ -0.266952$ & $ 0.680773 $ & $ -0.266944$ & $ 0.681278 $ & $ -0.266553$   \\ 
$-\frac{1}{2}$ &  $\frac{5}{2}$ & $2$ & $0.699601$   & $-0.452294$ & $0.699890$  &  $ -0.451648$ & $0.695105$  & $ -0.452281$ & $0.695394 $ & $ -0.451643$ & $ 0.650564 $ & $ -0.451653$ & $ 0.650861 $ & $ -0.451087$   \\ 
\hline 
\end{tabular}
\label{table8}
\end{table*}  
\endgroup

The first unexpected result concerns negative $qQ$ (electromagnetic attraction). Specifically, as $qQ$ becomes more negative, it has been recently found~\cite{konoplya,davi} that, for non-extremal black holes, there is a special point at which the real part of the fundamental quasinormal frequency becomes zero and its quasinormal branch is suddenly interrupted. In our numerical simulations for an extremal black hole, on the other hand, we have observed that the QNM branch does not disappear. This behaviour seems to be very general, occurring for the fundamental QNM and the first overtone of scalar and Dirac perturbations, as shown in Fig.~\ref{fig5}. A possible explanation for this difference is the fact that the analysis of convergence of the power series expansion in the continued fraction method breaks down when $\omega$ is a pure negative imaginary number, which is exactly what happens for near extremal black holes at the critical point.
In principle, this is also a problem for extremal black holes. However, as we increase the number of terms in the continued fraction truncations, the real part of the frequency of the QNMs seems to approach zero assymptotically as $qQ \rightarrow -\infty$.

The second unexpected result occurs for positive $qQ$ (electromagnetic repulsion). As $qQ$ increases, the quasinormal frequencies of a near extremal and an extremal black hole behave quite differently. This can be observed in Fig.~\ref{fig6} for the fundamental frequency and the first overtone of scalar and Dirac perturbations. One might argue that, since $Re(\omega) \rightarrow qQ$ as $qQ$ increases~\cite{hodqQ,konoplya,davi}, $J_0$ tends to become a positive real number and, therefore, convergence of the continued fraction method becomes compromised (as discussed in Sec.~\ref{cfm}). While it is true that convergence becomes slower as $qQ$ increases, by increasing the number of terms in the continued fractions, we were able to obtain accurate results for the range of parameters considered in this study.

\subsection{Mode stability of extremal black holes}

\begin{figure}
\begin{center}
\includegraphics[width=8.6cm]{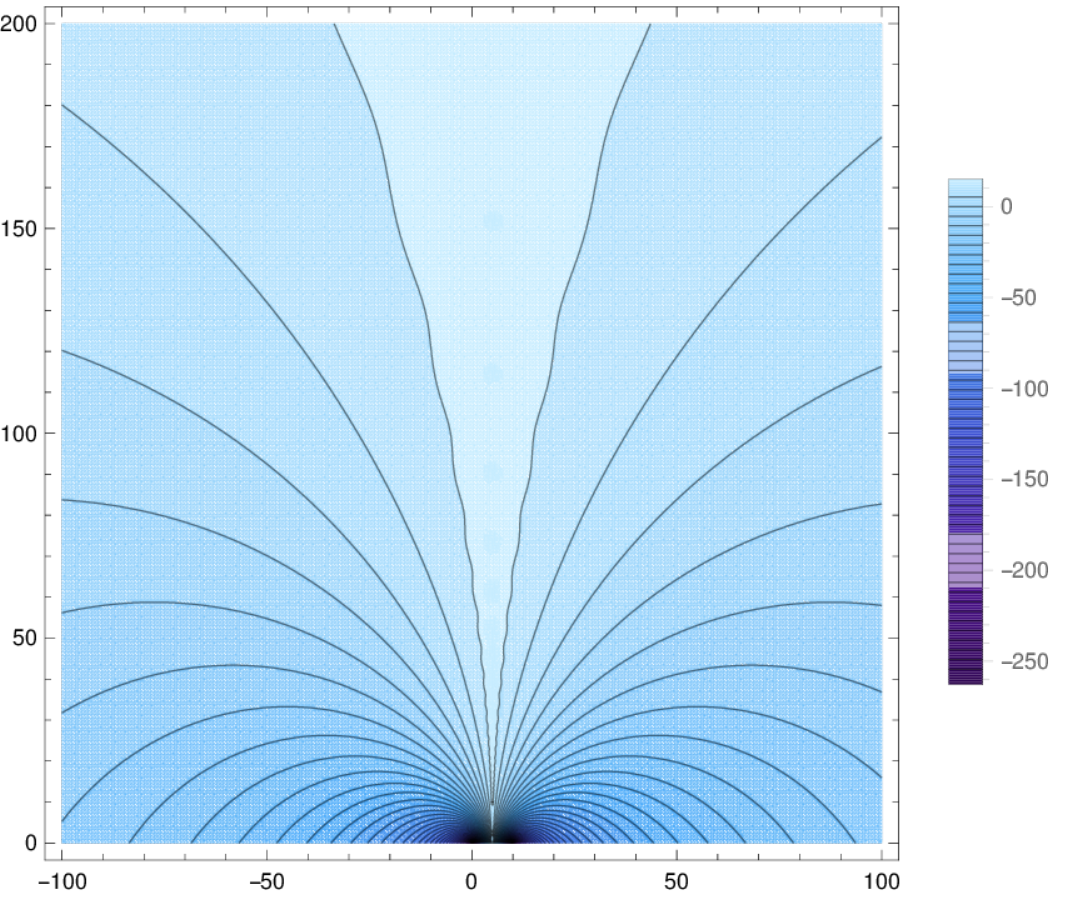}\\
\includegraphics[width=8.6cm]{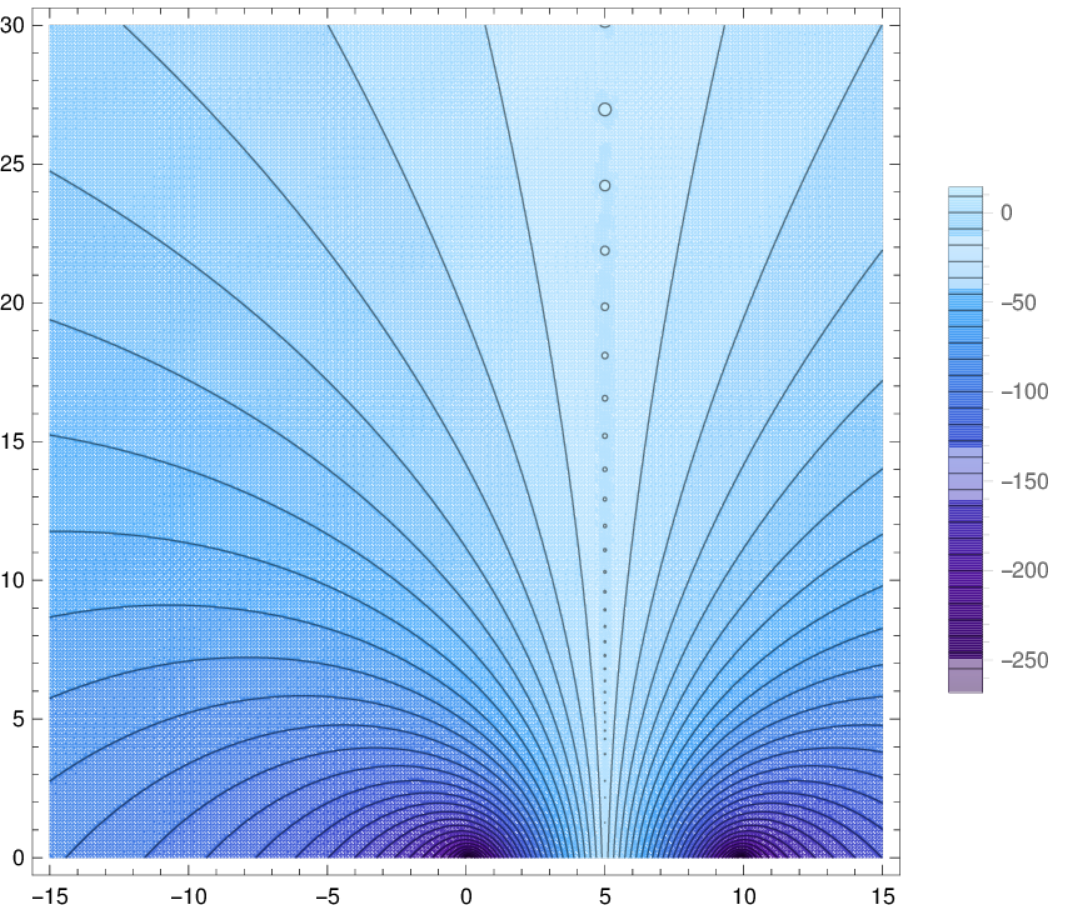}
\caption{(Colors online.) Searching for mode instabilities of the extremal RN black hole for $s=0$, $\ell=1$, and $qQ=10$: contour plots of the logarithm of the LHS of equation \eqref{cond1} as a function of $\operatorname{Re}(M\omega)$ and $\operatorname{Im}(M\omega)$. The number of terms in the continued fractions is 1000. The bottom plot is a close-up view of the upper one and suggests that unstable modes with small $\operatorname{Im}(M\omega)$ might exist. 
\label{fig7}
}
\end{center}
\end{figure}

\begin{figure*}
\begin{center}
\begin{tabular}{ccc}
\includegraphics[width=5.8cm]{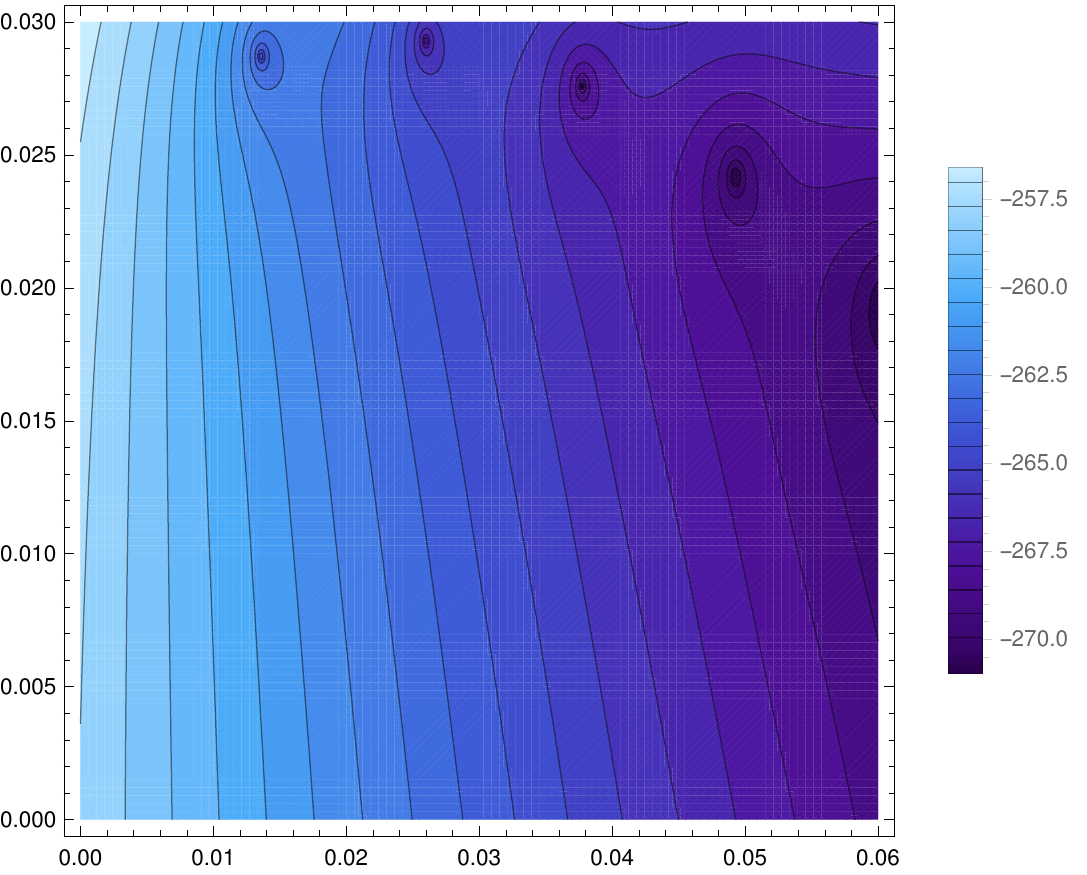}&
\includegraphics[width=5.8cm]{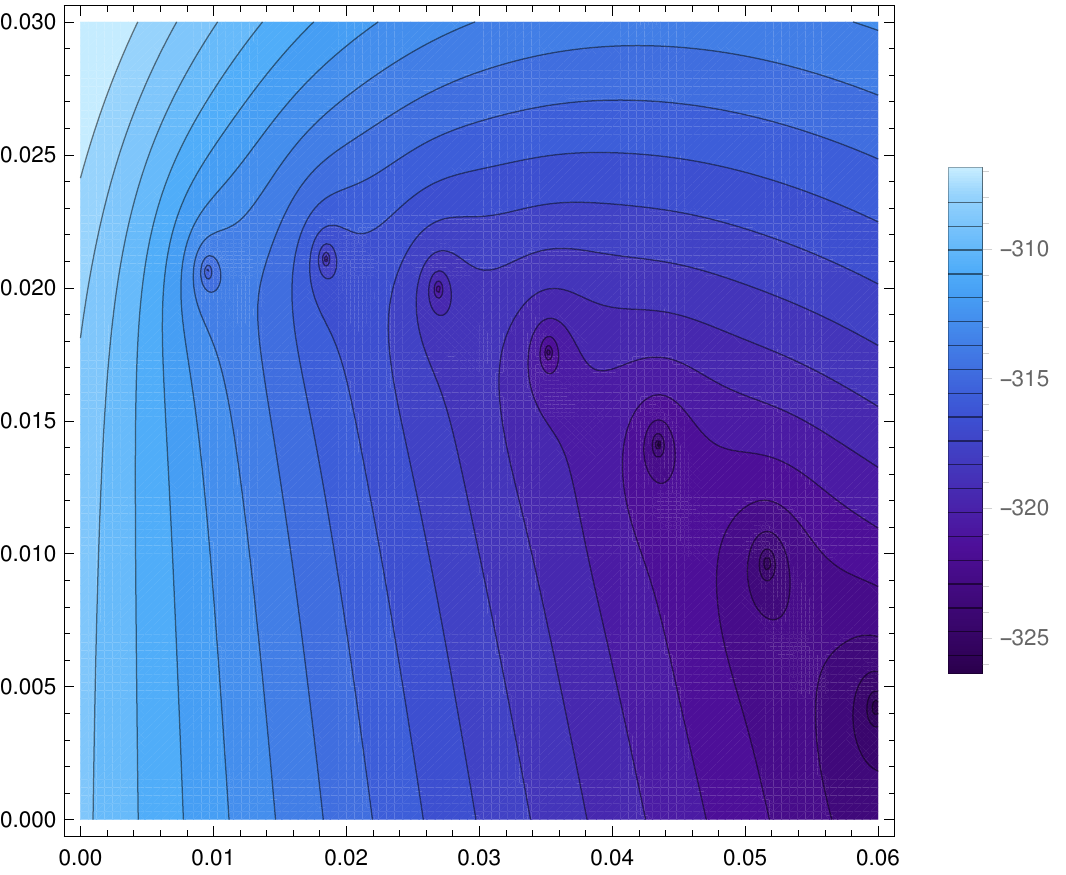}&
\includegraphics[width=5.8cm]{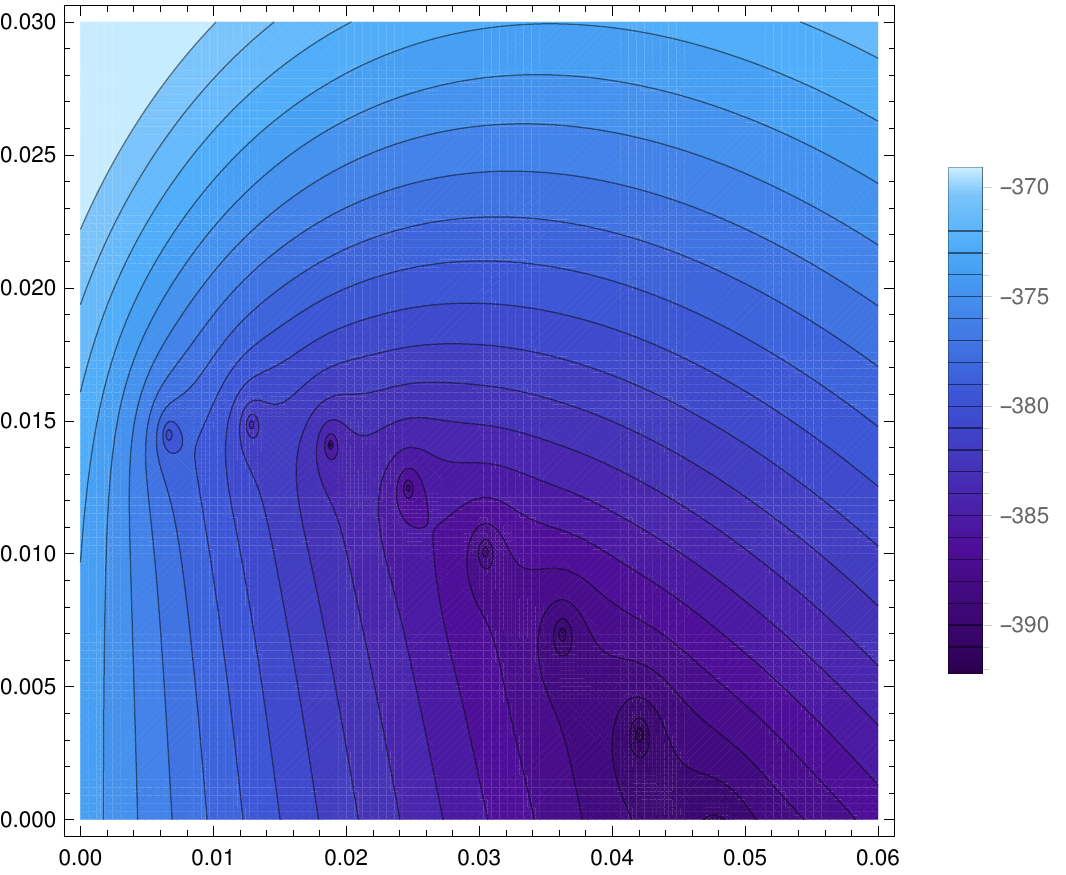}\\
\includegraphics[width=5.8cm]{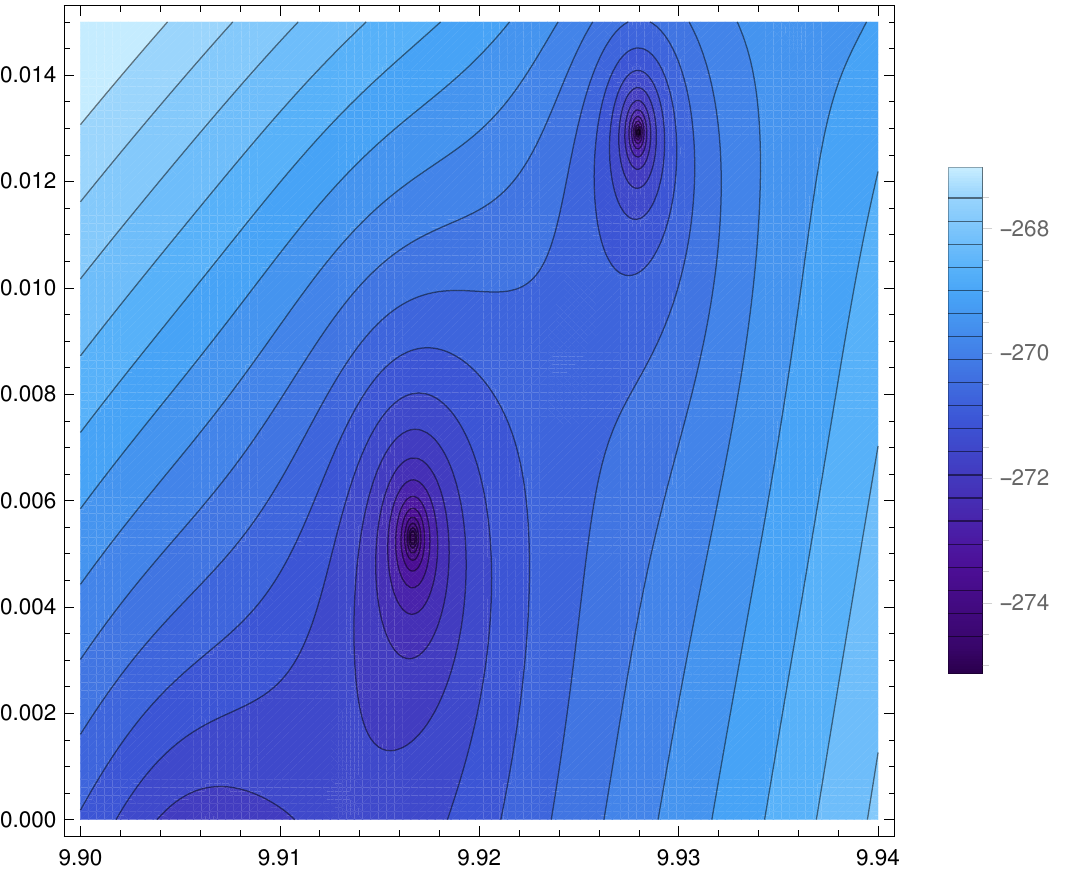}&
\includegraphics[width=5.8cm]{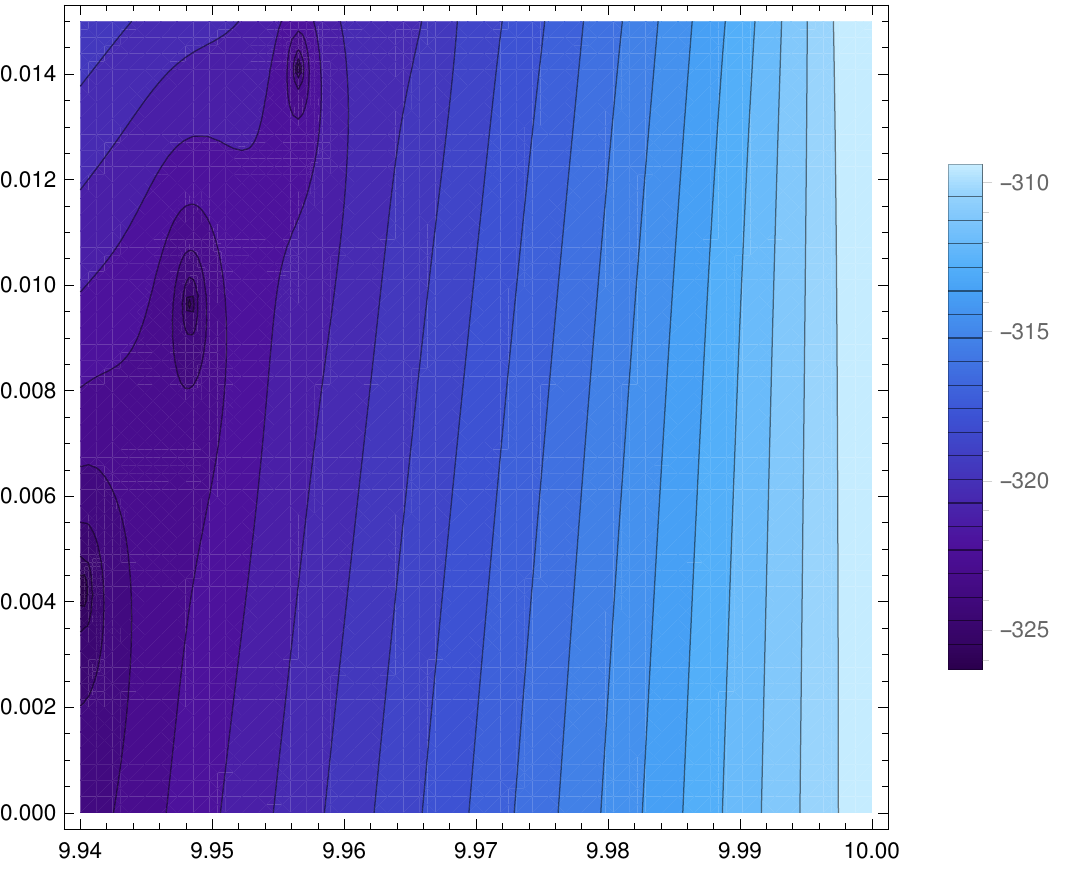}&
\includegraphics[width=5.8cm]{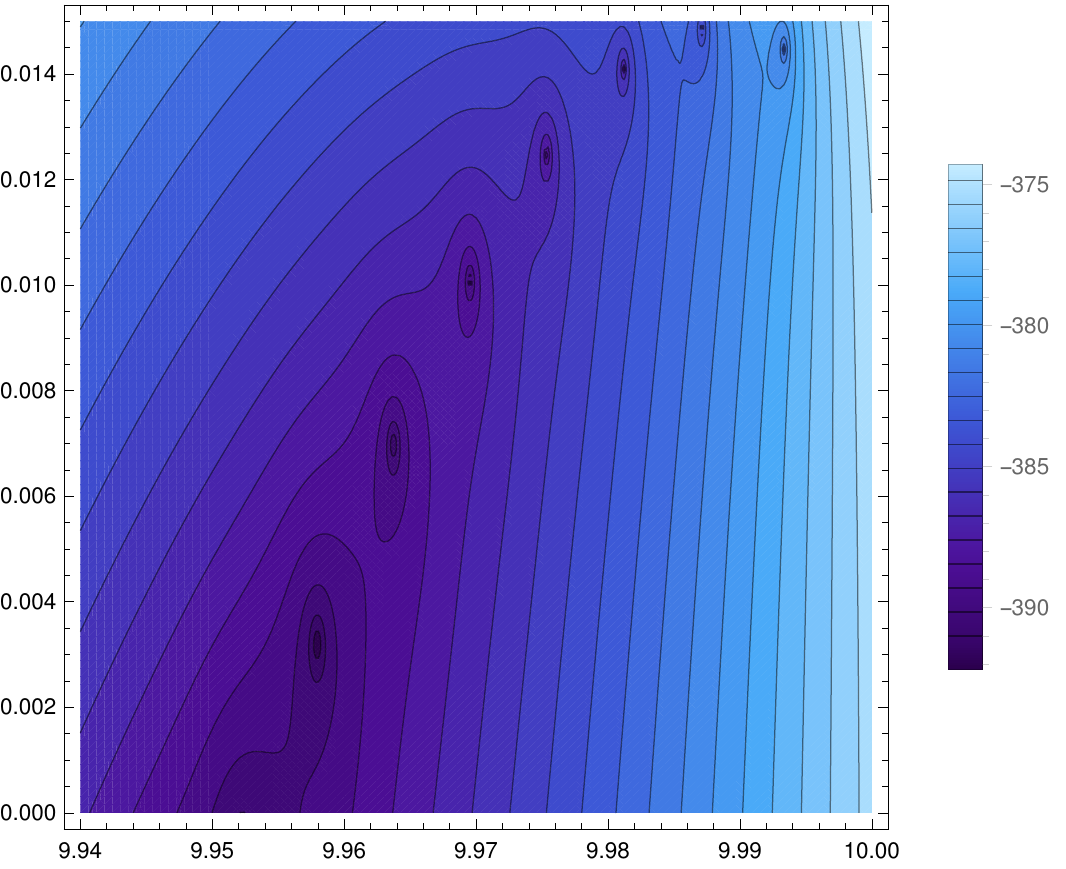}
\end{tabular}
\caption{(Colors online.) Searching for mode instabilities of the extremal RN black hole for $s=0$, $\ell=1$, and $qQ=10$: close-up view of the contour plots in Fig.~\ref{fig7} around $M\omega=0$ (top panels) and around $M\omega=qQ=10$ (bottom panels). From left to right, the panels correspond respectively to 1000, 1400 and 2000 terms in the continued fractions. As the number of terms is increased the minima move towards $\operatorname{Im}(M\omega)=0$, where we cannot guarantee converge of the continued fractions (as explained in Sec.~\ref{cfm}). This indicates that no unstable mode exists around $M\omega=0$ or $M\omega=10$.
\label{fig8}
}
\end{center}
\end{figure*}

\begin{figure*}
\begin{center}
\begin{tabular}{ccc}
\includegraphics[width=5.8cm]{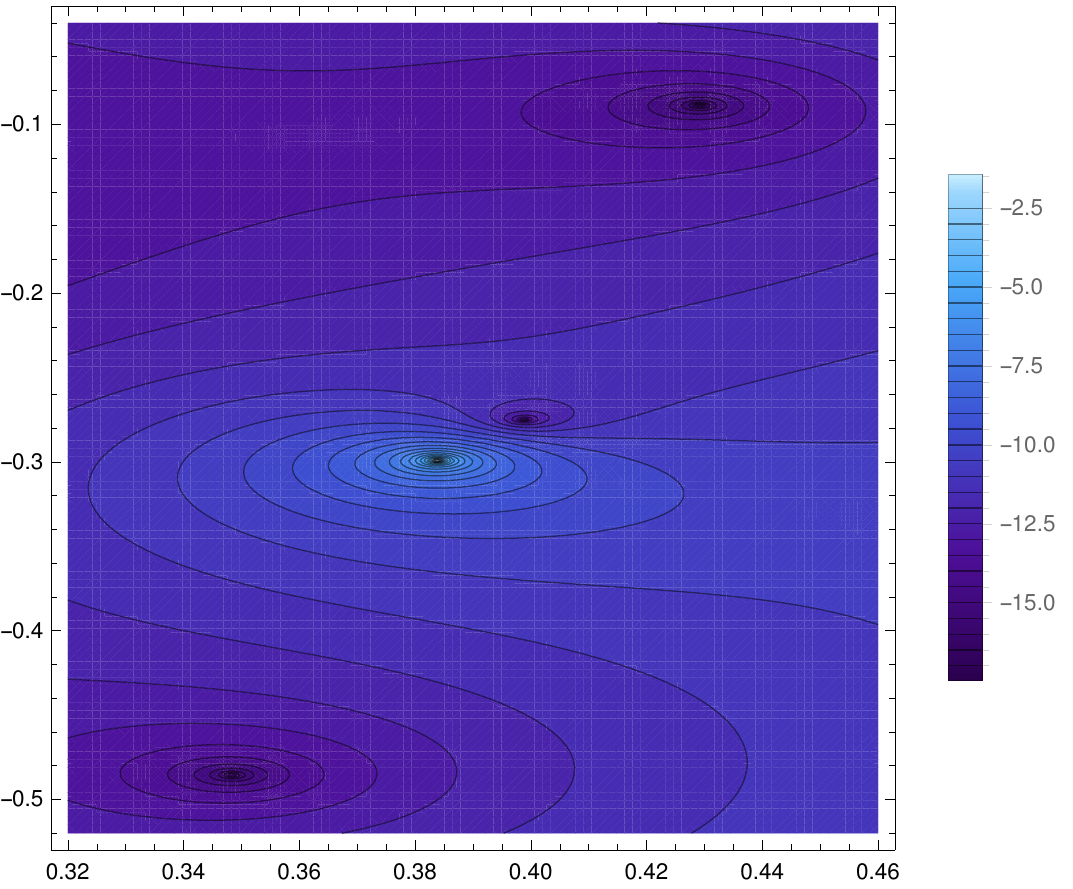}&
\includegraphics[width=5.8cm]{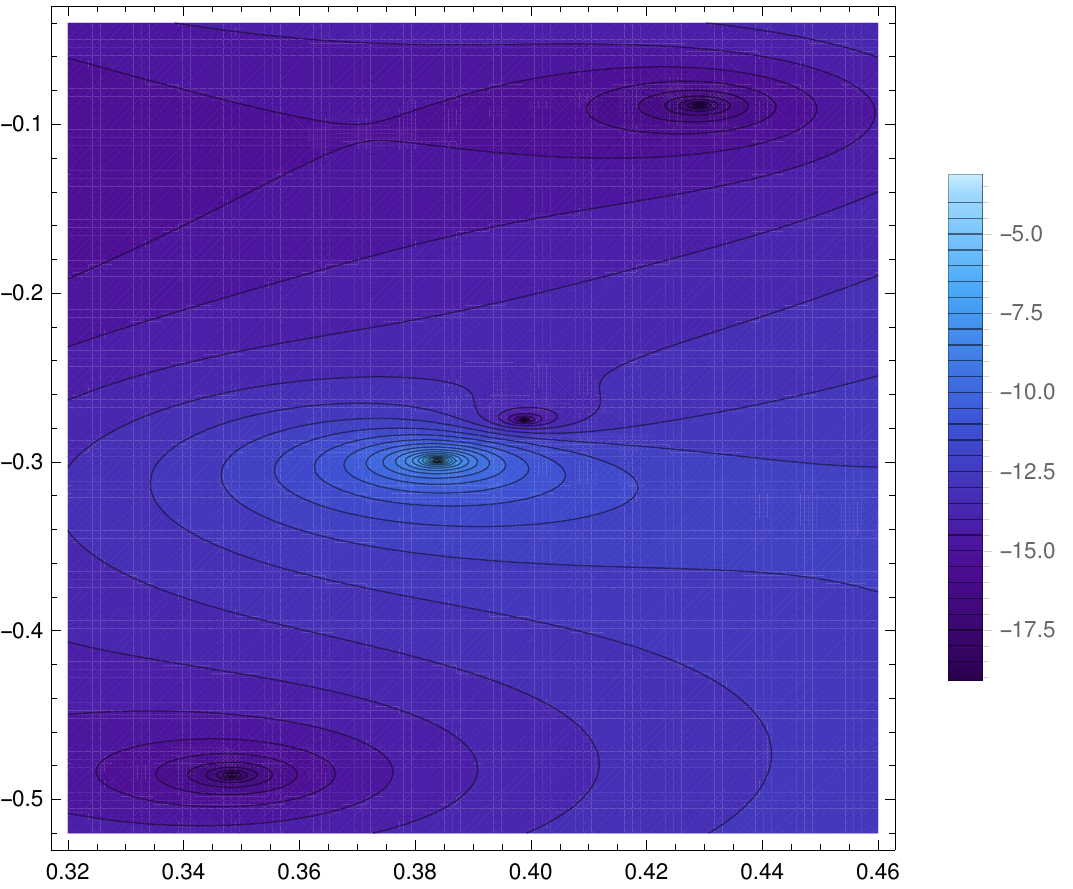}&
\includegraphics[width=5.8cm]{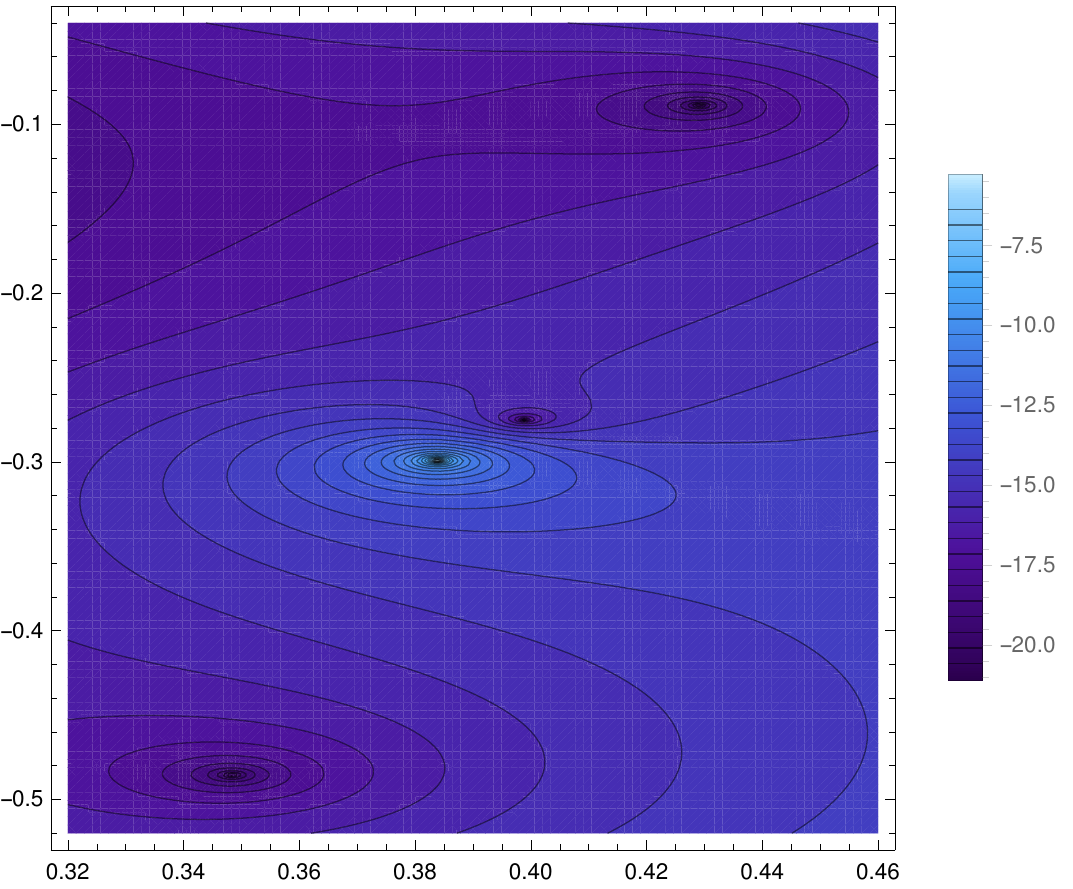}
\end{tabular}
\caption{(Colors online.) Contour plots indicating the location of the first three damped modes of the extremal RN black hole for $s=0$, $\ell=1$, and $qQ=0.1$. From left to right, the panels correspond respectively to 1000, 1400 and 2000 terms in the continued fractions. Unlike the contour plots in Fig.~\ref{fig8}, as the number of terms is increased, the minima do not move. The precise location of these quasinormal modes is calculated in Table \ref{table7}. We also note that the point located at $M\omega \approx 0.38-0.3i$ is a local maxima and, consequently, not a solution of \eqref{cond1}.
\label{fig9}
}
\end{center}
\end{figure*}

The final section is devoted to an analysis of the stability of extremal black holes. As explained in Ref.~\cite{herdeiroradu}, there exists three different notions of black hole stability: mode stability, linear stability and non-linear stability. The first one does not necessarily imply the second one, and the second one does not imply the third. Mode stability is related to the existence of QNMs with positive imaginary parts and, therefore, can be studied with the method described in this paper. 

Mode stability was established in Ref.~\cite{whiting} for non-extremal Kerr black holes and in Ref.~\cite{dias} for non-extremal Kerr-Newman black holes. In this work we address whether or not extremal Kerr and RN black holes are mode stable by searching for QNMs with positive imaginary parts. For extremal Kerr black holes, we start by solving, for several values of $s$, $\ell$ and $m$, the continued fraction equations \eqref{cond1} and \eqref{condang} (with 1000 terms) using as an initial guess for $M\omega$ different values with positive imaginary part [for $\lambda$ we use $(\ell-s)(\ell + s +1)$ as the initial guess]. If a solution $M\omega$ with positive imaginary part is found, the procedure is repeated with more terms in the continued fractions in order to confirm the results. For extremal RN black holes we proceed similarly, solving the continued fraction equation \eqref{cond1} for several values of $s$, $\ell$, and $qQ$.
Additionally, we also make contour plots of the logarithm of the LHS of equation \eqref{cond1} as a function of $\operatorname{Re}(M\omega)$ and $\operatorname{Im}(M\omega)$. If any unstable mode is suspected from such plots, we use the corresponding value of the frequency as an initial guess in the root finding algorithm to confirm the result. 

We report that no quasinormal frequencies with positive imaginary parts have been found (neither for extremal Kerr black holes nor for extremal RN black holes), indicating that these extremal black holes, like their non-extremal counterparts, are mode stable.

 In order to illustrate our search, we exhibit some of the plots obtained for scalar perturbations with $\ell=1$ and $qQ=10$ around an extremal RN black hole. In particular, note that the plots in Fig.~\ref{fig7} suggest that solutions with positive imaginary parts might exist around $M\omega = 0$ and around $M\omega=qQ=10$. However, as the number of terms in the continued fractions is increased, the minimum points of the contour plots move towards $\operatorname{Im}(M\omega)=0$ (see Fig.~\ref{fig8}). The same behaviour is found when attempting to solve equation \eqref{cond1}: as the number of terms in the continued fractions is increased, its solutions move towards the real axis, meaning that the continued fractions do not converge.

Conversely, the (stable) quasinormal modes found using our method are indeed convergent solutions. In Fig.~\ref{fig9} we illustrate this result graphically for the case $s=0$, $\ell=1$, $qQ=0.1$. One can see that, as the number of terms in the continued fractions is increased, the location of the maxima remain unchanged. As explained in Sec.~\ref{sec_4}, every time a solution was found, we increased the number of terms in equation \eqref{cond1} to confirm the result.

% 
%================
\section{Final remarks} \label{sec_5}
%================

In this paper, inspired by the ideas of Ref.~\cite{onozawa}, we have successfully implemented a continued fraction method to determine the QNMs of neutral massless perturbations around an extremal Kerr black hole and the QNMs of charged massless perturbations around an extremal RN black hole (we remark the unified framework employed, which is applicable in both situations). Starting with the perturbation equations~\eqref{teukoang} and \eqref{teukorad}, we set (exactly) $a=M$ or $Q=M$ and, using the asymptotic behaviour given in \eqref{bc1} and \eqref{bc2}, write a power series solution of the radial equation around the ordinary point $r=2M$.

 We were able to obtain the quasinormal frequencies for arbitrary values of the parameters $s$, $\ell$, $m$ and $qQ$. Our results agree with Leaver's original continued fraction method for near extremal black holes, as explicitly shown in tables \ref{table1},\ref{table6}-\ref{table8}. Our results also agree with previous results in the literature for near extremal Kerr black holes~\cite{leaver1,onozawa2,cardoso_data} and for near extremal RN black holes~\cite{konoplya, davi}. In particular, for $m \neq 0$, our numerical calculations, as shown in tables \ref{table2} and \ref{table5}, and in Fig.~\ref{fig1}, are compatible with the scalar and gravitational damped modes of near extremal Kerr black holes, which were studied in Refs.~\cite{branching1,branching2}. Besides extending the analysis of Refs.~\cite{branching1,branching2} to Dirac and electromagnetic fields, we have studied in detail the behaviour of the most stable damped modes as the parameters $\ell$ and $m$ are varied.  

It is important to note that the asymptotic behaviour of the solution of the radial perturbation equation of a near extremal black hole is not the same as \eqref{bc1}, which holds only for extremal black holes. The remarkable agreement that we have observed between the QNMs of near extremal and extremal black holes, for most sets of parameters $s, \ell, m$ and $qQ$, supports the idea that the QNM limit of extremal black holes is continuous.  Nevertheless, there are some important differences, as observed in Figs.~\ref{fig5} and \ref{fig6}. In particular, contrary to extremal RN black holes, near extremal black holes exhibit critical values of the electromagnetic parameter below which the quasinormal branches disappear. Moreover, as the electromagnetic interaction increases, notable differences between the quasinormal frequencies of extremal and near extremal RN black holes emerge.       

We would also like to remark that another commonly used method, the WKB method, besides being an approximation, becomes less accurate as the extremal limit is approached~\cite{wkbkerr} (see also Ref.~\cite{branching2} for an implementation which works for near extremal black holes, but only in the eikonal limit). Since the original continued fraction method fails when $a=M$ or $Q=M$, our implementation of the ideas of Ref.~\cite{onozawa} in such a case provides the most accurate method to determine the QNMs of an extremal Kerr black hole and the QNMs of charged perturbations around an extremal RN black hole. 

Finally, we point out that we have found no quasinormal frequencies which possess positive imaginary parts, indicating that extremal Kerr and extremal RN black holes are mode stable. Onozawa et.~al, similarly, did not report finding unstable modes related to (uncharged) perturbations around an extremal RN black hole~\cite{onozawa}. Recently, however, extremal RN and Kerr black holes have been shown to be linearly unstable. The first results were obtained by Aretakis in Refs.~\cite{aretakis1,aretakis2,aretakis3,aretakis4}. Later, inspired by Aretakis' works, Lucietti and Reall~\cite{lucietti} have shown that extremal Kerr black holes are unstable under linearized gravitational and electromagnetic perturbations. There might be, however a connection between zero-damping modes and these linear instabilities. While for near extremal black holes the zero-damping modes have small (but non-zero) negative imaginary parts, it is believed that, for extremal black holes, they are purely real~\cite{ferrari,glampe,cardosox,hod00}. There has been some suggestions that these purely real modes are related to instabilities~\cite{detweiler,davi}, but whether this is really the case (and their relation with these linear instabilities) has yet to be established.     
  
%================
\acknowledgments
%================
The author is grateful to V. Cardoso, D. Giugno, J. P. Pitelli, and A. Saa for enlightening discussions and feedback. The author is also grateful to S. Weinfurtner and the University of Nottingham for hospitality while this work was being completed. M.~R.~was partially funded by the S\~ao Paulo Research Foundation (FAPESP), Grants No. 2013/09357-9 and No. 2015/14077-0.

%================
\appendix*
%================

\section{Recurrence relation, gaussian elimination and decoupling }

The coefficients for the 5-term recurrence relation~\eqref{5term} are given explicitly by 
\begin{align}
\alpha_n &= n^2 + n, \quad \beta_n = P_1 n, \\
 \gamma_n &= -2n^2 + P_2 n + P_3, \quad \delta_n = P_4 n + P_5, \\
\epsilon_n &= n^2 + P_6 n + P_7,
\end{align}
where
 \begin{align}
P_1 &= 4 (im -iM\omega - s), \\
P_2 &=2(1-4im+16iM\omega), \\
P_3 &=  -2 [1 + 2 s + 2 \lambda + 2 iM \omega (4  +  23i M \omega) \\
 & \ \ \ -2 im (1 - 6i M \omega)], \\
  P_4 &= 4(im - i M\omega + s), \\
 P_5 & = -4 (im - iM\omega + s)(1 + 4iM\omega), \\
 P_6 & = -3 - 8iM\omega, \\
 P_7 &= 2 [1 + 6iM\omega - 8 (M\omega)^2],
\end{align}
for an extremal Kerr black hole, and
 \begin{align}
P_1 &=  4 (iqQ - s), \\
P_2 &=2(1-6iqQ+12iM\omega), \\
P_3 &=  -2 [1 - 3 i qQ + 2 s + 2 \lambda + 6 iM\omega  \\
 & \ \ \ -  2(qQ - 4 M \omega) (3 qQ - 4 M \omega)], \\
 P_4 &= 4(iqQ + s), \\
 P_5 & = -4 (iqQ + s)(1 - 2iqQ + 4iM\omega), \\
 P_6 & = -3 + 4iqQ - 8iM\omega, \\
 P_7 &=  2 (1-2iqQ + 4iM\omega)(1 - iqQ + 2iM\omega),
\end{align}
for an extremal RN black hole.
In order to reduce the 5-term recurrence relation~\eqref{5term} to a 3-term recurrence relation, we use a double gaussian elimination procedure~\cite{leaver2,onozawa}. We define 
\be \label{gauss1}
\begin{cases}
\alpha_1' = \alpha_1, \ \beta_1' = \beta_1, \ \gamma_1' = \gamma_1,  \\
\alpha_2' = \alpha_2, \ \beta_2' = \beta_2, \ \gamma_2' = \gamma_2,  \ \delta_2' = \delta_2 \\
\epsilon_n' = 0, \ \alpha_n' = \alpha_n, \ \beta_n' = \beta_n - \frac{\epsilon_n}{\delta_{n-1}'} \gamma_{n-1}', \qquad n \ge 3\\
 \gamma_n' = \gamma_n - \frac{\epsilon_n}{\delta_{n-1}'} \beta_{n-1}', \delta_n' = \delta_n - \frac{\epsilon_n}{\delta_{n-1}'} \delta_{n-1}' , \qquad n \ge 3 \\
\end{cases}
\ee 
followed by
\be \label{gauss2}
\begin{cases}
\alpha_1'' = \alpha_1', \ \beta_1'' = \beta_1', \ \gamma_1'' = \gamma_1',  \\
\alpha_n'' = \alpha_n', \ \beta_n'' = \beta_n' - \frac{\delta_n'}{\gamma_{n-1}''} \alpha_{n-1}'' , \qquad n \ge 2 \\ \gamma_n'' = \gamma_n' - \frac{\delta_n'}{\gamma_{n-1}''} \beta_{n-1}'', 
 \delta_n'' = 0 , \qquad n \ge 2 \\
\end{cases}
\ee 
transforming eqs.~\eqref{rec0}-\eqref{5term} into
\be \label{3term2}
\alpha_n '' a_{n+1} + \beta_n '' a_n + \gamma_n '' a_{n-1}=0, \qquad n \ge 1.
\ee

This 3-term recurrence relation couples even ($a_{2n}$) and odd ($a_{2n+1}$) expansion coefficients. In order to decouple them, we proceed as follows. First, we write \eqref{3term2} for $2n+1$ and solve for $a_{2n+2}$. Next, we write \eqref{3term2} for $2n$, solve for $a_{2n}$, and plug the result into the expression obtained earlier for $a_{2n+2}$, resulting in
\be 
a_{2n+2} = - \frac{\beta_{2n+1}''}{\alpha_{2n+1}''}a_{2n+1} + \frac{\gamma_{2n+1}''(\alpha_{2n}''a_{2n+1} + \gamma_{2n}''a_{2n-1} )}{\alpha_{2n+1}''\beta_{2n}''}. 
\ee  

Finally, we substitute this expression into \eqref{3term2} for $2n+2$, successfully decoupling the odd terms from the even: 
\be 
\alpha_n ^o d_{n+1} + \beta_n ^o d_n + \gamma_n ^o  d_{n-1} = 0, \qquad n \ge 1,
\ee
where $d_n=a_{2n+1}$ and the recurrence coefficients are given by
\be
\begin{cases} \label{odd}
 \alpha_n ^o & = \alpha_{2n+2}'', \\
\beta_n ^o &= \gamma_{2n+2}'' - \frac{\beta_{2n+1}'' \beta_{2n+2}''}{\alpha_{2n+1}''} +  \frac{\beta_{2n+2}'' \alpha_{2n}'' \gamma_{2n+1}''}{\alpha_{2n+1}'' \beta_{2n}''}, \\
 \gamma_n ^o & = \frac{\beta_{2n+2}'' \gamma_{2n}'' \gamma_{2n+1}''}{\alpha_{2n+1}'' \beta_{2n}''}.
\end{cases}
\ee

Analogously, it is possible to write a 3-term recurrence relation for the even terms:
\be 
\alpha_n ^e c_{n+1} + \beta_n ^e c_n + \gamma_n ^e c_{n-1} = 0, \qquad n \ge 1,
\ee
where $c_n=a_{2n}$ and 
\be
\begin{cases} \label{even}
 \alpha_n ^e & = \alpha_{2n+1}'', \\
 \beta_n ^e &= \gamma_{2n+1}'' - \frac{\beta_{2n}'' \beta_{2n+1}''}{\alpha_{2n}''} +  \frac{\beta_{2n+1}'' \alpha_{2n-1}'' \gamma_{2n}''}{\alpha_{2n}'' \beta_{2n-1}''}, \\
 \gamma_n ^e & = \frac{\beta_{2n+1}'' \gamma_{2n-1}'' \gamma_{2n}''}{\alpha_{2n}'' \beta_{2n-1}''}.
\end{cases}
\ee

Note that, while the original decoupling procedure of Ref.~\cite{onozawa} only works when $\delta_n = 0$, our decoupling works for $\delta_n \neq 0$ (which is the case in all scenarios we have analysed in this paper).

%================

%\bibliographystyle{unsrt}
\bibliographystyle{apsrev}
\bibliography{qn_bib2}

\end{document}